\documentclass[lettersize,journal]{IEEEtran}
\pdfoutput=1
\usepackage{amsmath,amsfonts}
\usepackage{algorithmic}
\usepackage{array}
\usepackage[caption=false,font=normalsize,labelfont=sf,textfont=sf]{subfig}
\usepackage{textcomp}
\usepackage{stfloats}
\usepackage{url}
\usepackage{verbatim}
\usepackage{graphicx}
\usepackage{cite}
\usepackage{amsmath}
\usepackage{amssymb}
\usepackage{graphicx}
\usepackage{epstopdf}
\usepackage{multirow}
\usepackage[linesnumbered,ruled]{algorithm2e}

\usepackage{graphicx}
\usepackage{subfig}
\def\BibTeX{{\rm B\kern-.05em{\sc i\kern-.025em b}\kern-.08em
    T\kern-.1667em\lower.7ex\hbox{E}\kern-.125emX}}
\usepackage{balance}
\begin{document}

\title{Construction of PAC Codes with List-Search and Path-Splitting Critical Sets}

\author{Shirong Jiang, Jiahao Wang, Chunfang Xia and Xiang Li
\thanks{}
\thanks{The authors are with the School of Mechanical Engineering and Electronic Information, China University of Geosciences, Wuhan 430074, China (e-mail: 1589178389@qq.com; jiahaow@cug.edu.cn; 1224350759@qq.com; lix@cug.edu.cn).}
\thanks{}
}

\markboth{}%
{Construction of PAC Codes with List-Search and Path-Splitting Critical Sets}

\maketitle

\begin{abstract}
Polarization-adjusted convolutional (PAC) codes can approach the theoretical bound for block error rate (BLER) performance at short-to-medium codeword length. 
PAC codes have excellent BLER performance using Monte Carlo (MC) rate-profiles and Weighted Sum (WS) rate-profiles, but the BLER performances of the constructed codes still fall away from the dispersion bound at high signal-to-noise ratios (SNR). 
This paper proposes a List-Search (LS) construction method for PAC codes, which considers the influence of weight spectrum on BLER performance and the condition that sequence decoding for PAC codes having a finite mean computational complexity. 
The proposed construction method using LS can reduce the number of minimum weight codewords of PAC codes. 
The BLER performance of the constructed codes is better than that of the constructed codes using MC rate-profiles or WS rate-profiles, and can approach the dispersion bound at high SNR. 
Moreover, the BLER performance of successive cancellation list (SCL) decoding PAC codes using LS rate-profiles can approach the theoretical bound, but SCL decoding requires a large number of sorting operations. 
To reduce the number of sorting operations, a path-splitting critical sets (PSCS) construction method is proposed. 
The PSCS obtained by this method are the information bits subset that have the greatest influence on the number of minimum weight codewords. 
The simulation results show that this method can significantly reduce the number of sorting operations during SCL-type decoding.
\end{abstract}


\section{Introduction}
\IEEEPARstart{P}{olar} codes \cite{ref1} are a class of linear block codes proposed by Arıkan based on channel polarization phenomenon. 
The capacity of any binary-input discrete memoryless channel (BI-DMC) can be realized when the code length goes to infinity by successive cancellation (SC) decoding. 
However, for finite (small to medium) block lengths, the block error rate (BLER) performance of SC decoding is poor compared with the state-of-the-art low-density parity-check (LDPC) codes and Turbo codes. 
To improve the BLER performance of polar codes with finite block length, both SC list (SCL) decoding \cite{ref2} and SC stack (SCS) decoding \cite{ref3} have been proposed successively, whose BLER performance is close to that of maximum likelihood (ML) decoding. 
In addition, the BLER performance of CRC-Aided SCL(CA-SCL) decoding \cite{ref2}, \cite{ref4} and CRC-Aided SCS(CA-SCS) decoding \cite{ref4} can compete with the state-of-the-art LDPC codes and Turbo codes with the help of cyclic redundancy check (CRC) codes.

With finite block lengths, although the polar codes can obtain excellent BLER performance using CA-SCL or CA-SCS decoding, it still cannot reach the theoretical bound. 
Because channel polarization occurs relatively slowly with finite block lengths, the 0-1 rate assignments of polar codes waste capacity of the bit channel. 
Recently, Arikan used Pinsker scheme \cite{ref5} to propose the concatenation code of rate-1 convolution transform and polar transform, namely polarization-adjusted convolutional (PAC) codes \cite{ref6}. 
PAC codes can prevent capacity loss and improve weight distribution, and BLER performance is close to the dispersion approximation (or dispersion bound) \cite{ref9} using sequential decoding \cite{ref6} or list decoding \cite{ref7}, \cite{ref8}. 

Although PAC codes have excellent BLER performance, their BLER performance is very dependent on the rate-profiles construction. 
When Arıkan proposed PAC codes, he gave polar rate-profiles and Reed-Muller(RM) rate-profiles, which showed different BLER performance and computational complexity using sequence decoding \cite{ref6}. 
The polar rate-profiles use Gaussian approximation (GA) rate-profiles, and information bits are selected based on the reliability of sub-channels \cite{ref10}. 
RM rate-profiles can make PAC code close to the dispersion bound, but requires higher computational complexity. 
Although the RM rate-profiles has excellent BLER performance, it is only applicable to specific code parameters. 
For a wider range of code parameters, RM-polar rate-profiles was obtained by combining the construction of RM codes and polar codes \cite{ref11}. 
The RM-polar rate-profiles not only considers the row weight of the polar generation matrix, but also considers the reliability of the sub-channels, so a larger minimum hamming weights can be obtained. 

Under a wider range of code parameters, although the construction codes obtained from the RM-polar rate-profiles can have lower computational complexity, its BLER performance deviates significantly from the dispersion bound. 
To apply PAC codes to any code length and code rate, Monte Carlo (MC) rate-profiles is proposed \cite{ref12}. 
While the MC rate-profiles improves BLER performance, the MC rate-profiles requires constant iteration that is empirical and complex for construction. 
In addition, considering the cutoff rate and the utilizations of the polarized sub-channels, the weighted sum (WS) rate-profiles are proposed \cite{ref13}. 
The WS rate-profiles have  slightly better BLER performance than the MC rate-profiles. 
Although the MC rate-profiles and WS rate-profiles can obtain better BLER performance than the RM-polar rate-profiles, the BLER performance still deviates from the dispersion bound at high SNRs. 
To make the constructed code performance close to the dispersion bound, according to genetic algorithm based construction of polar codes \cite{ref14}, a modified genetic algorithm based construction of PAC codes was proposed \cite{ref15}. 
This method can reduce the number of minimum weight codewords of PAC codes and achieve BLER performance close to the dispersion bound. 
However, the construction of PAC codes with modified genetic algorithm has high design complexity, and needs constant iteration. 

Using RM construction, the BLER performance of PAC codes List (or SCL) decoding with a code length of 128 and code rate of $1/2$ can approach the dispersion bound \cite{ref7}, \cite{ref8}. 
However, SCL decoding for PAC codes has the high computational complexity and a large number of sorting operations. 
Pruned SCL decoding \cite{ref16} is used to prune unnecessary paths after sorting operation during SCL decoding. 
Pruned SCL decoding can significantly reduce the computational complexity and number of sorting operations. 
Moreover, an alternative SCL decoding pruning technique using a stack is proposed in \cite{ref17}. 
The pruning techniques mentioned above are carried out after sort operation. 
To further reduce the computational complexity and number of sorting operation, SCL decoding with modified pruning technology (PSCL) was proposed \cite{ref18}. 
Although pruning technology can significantly reduce the computational complexity and the number of sorting operation during SCL decoding, it varies with the change of SNR. 

This paper proposes a construction method for PAC codes with List-Search (LS) that can be applied to any PAC codes with any code length and rate. 
The BLER performance of the constructed codes is close to the theoretical bound. 
LS construction are designed by using LS metric, and the final search results include codes with weight maximization and the number of minimum weight codewords. 
When the rates coincide, the proposed LS construction and RM construction can construct the same codes, so LS construction can be regarded as a generalization of RM construction. 
In addition, LS construction takes into account the finite average computational complexity of sequence decoding for PAC codes, so that the construted code can be tradeoff between decoding performance and complexity. 
To further reduce the number of sort operations during SCL-type decoding, a construction method with path-splitting critical sets (PSCS) is proposed using RM rate-profiles (or generalization of RM rate-profiles). 
PSCS can be obtained by PSCS construction, while path-splitting and sort operations occur only at PSCS during list decoding, which can greatly reduce the number of sort operations. 
The numerical results show that the proposed PSCS construction greatly reduces the number of  sort operations during SCL decoding, Pruned SCL decoding and PSCL decoding while maintaining the same BLER performance. 

The remainder of this article is structured as follows. 
Section II briefly reviews the channel polarization phenomenon and PAC codes, and describes list decoding and Fano decoding. 
In Section III, the influence of the number of minimum weight codewords on the BLER performance of ML decoding is first analyzed. 
Then the conditions of finite average computational complexity of sequence decoding for PAC codes are summarized. 
Finally, the LS construction method for PAC codes is proposed. 
In Section IV, the effect of weight spectrum on BLER performance is considered, and the complete path-splitting critical sets (CPSCS) is obtained and the PSCS construction method is proposed. 
In Section V, the simulation results of LS construction and PSCS construction for PAC codes are obtained. 
Finally, Section VI provides a succinct overview of this work.

\section{Preliminaries}
\subsection{Channel Polarization}
Channel polarization means that $N$ independent copies of a binary-input discrete memoryless channel (B-DMC) form $N$ new bit-channels $\{{W}^{(i)}_{N}:0\leqslant i < N\}$ \cite{ref1} through channel combining and channel splitting. 
With $N\rightarrow + \infty$, the channel capacity $I({W}^{(i)}_{N})$ of the channel approach either 0 or 1. 

Let ${W:\mathcal{X}\rightarrow\mathcal{Y}}$ denote a B-DMC channel with the input alphabet $\mathcal{X}=\{0,1\}$ and the output alphabet $\mathcal{Y}$ as a real number. 
The channel transition probability is defined by $W(\mathcal{Y}|\mathcal{X})$, $x\in\mathcal{X}$, $y\in\mathcal{Y}$.
Channel combining means that vector channel ${W}^{N}$ creates a new vector channel ${W}_{N}:{\mathcal{U}}^{N}\rightarrow{\mathcal{Y}}^{N}$ by mapping ${G}_{N}:{\{0,1\}}^{N}\rightarrow{\{0,1\}}^{N}$, where $\mathcal{U}=\{0,1\}$, ${G}_{N}={\mathbf{F}}^{\otimes n}=\scriptsize{\begin{bmatrix} {\mathbf{F}}^{\otimes(n-1)} & \mathbf{0} \\ {\mathbf{F}}^{\otimes(n-1)} & {\mathbf{F}}^{\otimes(n-1)} \end{bmatrix}}$, ${\mathbf{F}}^{\otimes 1}=\scriptsize{\begin{bmatrix} 1 & 0 \\ 1 & 1 \end{bmatrix}}$, $n={\log}_{2}{N}$. 
Channel splitting means that ${W}^{N}$ is split into $N$ binary-input coordinate channels ${W}^{(i)}_{N}:\mathcal{U}\rightarrow{\mathcal{Y}}^{N} \times {\mathcal{U}}^{i-1}$,$0\leqslant i < N$. 

When $\mathcal{X}$ is equal probability distribution $\{0,1\}$, channel capacity $I(W)$, which is defined as 
\begin{equation}
\label{ex1}
I(W) \triangleq \sum_{y \in \mathcal{Y}}\sum_{x \in \mathcal{X}} {\frac{1}{2}W(y|x)} {{\log}\frac{W(y|x)}{\frac{1}{2}W(y|0)+\frac{1}{2}W(y|1)}}.
\end{equation}
channel capacity $I(W)$ represents the maximum code rate for transmission over the channel $W$. 

One important parameter of a channel is the Bhattacharyya parameter $Z(W)$, which is defined as
\begin{equation}
\label{ex2}
Z(W) \triangleq \sum_{y \in \mathcal{Y}} {\sqrt{W(y|0)W(y|1)}}.
\end{equation}
Bhattacharyya parameter $Z(W)$ represents the upper bound of ML decoding error probability. 

Another important parameter in the channel is the cutoff rate ${E}_{0}{(1,W)}$ is 
\begin{equation}
\label{ex3}
{E}_{0}{(1,W)} = {1-{\log}_{2}{(1+Z(W))}}.
\end{equation}

After channel polarization, Bhattacharyya parameter $Z({W}^{(i)}_{N})$ and cutoff rate ${E}_{0}{(1,{W}^{(i)}_{N})}$ of bit channel ${W}^{(i)}_{N}$ are respectively
\begin{equation}
\label{ex4}
Z({W}^{(i)}_{N}) \triangleq \sum_{\mathbf{y} \in {\mathcal{Y}}^{N}} \sum_{{u}^{i-1}_{0} \in {\mathcal{U}}^{i-1}} {\sqrt{{W}^{(i)}_{N}(\mathbf{y},{u}^{i-1}_{0}|0){W}^{(i)}_{N}(\mathbf{y},{u}^{i-1}_{0}|1))}}.
\end{equation}
\begin{equation}
\label{ex5}
{E}_{0}{(1,{W}^{(i)}_{N})} = {1-{\log}_{2}{(1+Z({W}^{(i)}_{N}))}}.
\end{equation}

In the additive white Gaussian noise (AWGN) channel, the estimated value of Bhattacharyya parameter of bit channel ${W}^{(i)}_{N}$ \cite{ref19} is
\begin{equation}
\label{ex6}
Z({W}^{(i)}_{N}) = {\exp}^{-\frac{1}{2{({\sigma}^{(i)}_{N})}^{2}}}.
\end{equation}
The noise variance ${({\sigma}^{(i)}_{N})}^{2}$ of bit channel ${W}^{(i)}_{N}$ can be given by 
\begin{equation}
\label{ex7}
{({\sigma}^{(i)}_{N})}^{2} = \frac{2}{{m}^{(i)}_{N}}.
\end{equation}
Where, ${{m}^{(i)}_{N}}$ denote mean of the log-likelihood ratio (LLR) density function of bit channel ${W}^{(i)}_{N}$ \cite{ref19}.

\subsection{PAC Codes}

A PAC codes is characterized by $(N,K,\mathcal{A},\mathbf{g})$, where $N={2}^{n}$, $n\geqslant0$ is the block length, $K$ is the number of information bits, $\mathcal{A}\subset\{0,1,\cdots,N-1\}$ represents the set of $K$ information bit positions, and $\mathbf{g}$ is the convolutional transform polynomial coefficients vector. 
The complement of $\mathcal{A}$ is the index sets ${\mathcal{A}}^{c}$ with $N-K$ frozen bit. 
If the information index sets $\mathcal{A}$ and the convolutional transform polynomial coefficients vector $\mathbf{g}$ are fixed, PAC codes is represented by PAC($N,K$). 

The rate profiling in Fig. 1 representation embeds the information bit vector $\mathbf{d}={d}^{K-1}_{0}=\{{d}_{0},\cdots,{d}_{K-1}\}$ in vector $\mathbf{v}={v}^{N-1}_{0}=\{{v}_{0},\cdots,{v}_{N-1}\}$ so that $\mathbf{v}_{\mathcal{A}}=\mathbf{d}$ and $\mathbf{v}_{\mathcal{A}^{c}}=\mathbf{0}$. 
The rate profiling of PAC codes is similar to that of polar codes, transmitting useful information in the information bit and known information in the frozen bit. 
If the convolutional transform polynomial coefficient vector $\mathbf{g}=\{{g}_{0},\cdots,{g}_{m}\}$, ${g}_{0}={g}_{m}=1$,where $m+1$ is the constraint length of the convolution code, then the convolutional codewords $\mathbf{u}={u}^{N-1}_{0}=\{{u}_{0},\cdots,{u}_{N-1}\}$ can be obtained
\begin{equation}
\label{ex8}
{u}_{i}=\sum^{m}_{j=0} {{g}_{j}{v}_{i-j}}.
\end{equation}
Similarly, the matrix form of the convolution transform is $\mathbf{u}=\mathbf{v}\mathbf{T}$, where $\mathbf{T}$ is an upper-triangular Toeplitz matrix, 
\begin{equation}
\label{ex9}
\mathbf{T}=
\begin{bmatrix}
{g}_{0} & {g}_{1} & {g}_{2} & \cdots & {g}_{m} & 0 & \cdots & 0 \\
0 & {g}_{0} & {g}_{1} & {g}_{2} & \cdots & {g}_{m} & \ & \vdots \\
0 & 0 & {g}_{0} & {g}_{1} & \ddots & \cdots & {g}_{m} & \vdots \\
\vdots & 0 & \ddots & \ddots & \ddots & \ddots & \ & \vdots \\
\vdots & \ & \ddots & \ddots & \ddots & \ddots & {g}_{2} & \vdots \\
\vdots & \ & \ & \ddots & 0 & {g}_{0} & {g}_{1} & {g}_{2} \\ 
\vdots & \ & \ & \ & 0 & 0 & {g}_{0} & {g}_{1} \\ 
0 & \cdots & \cdots & \cdots & \cdots & 0 & 0 & {g}_{0} \\
\end{bmatrix}
.
\end{equation}
\begin{figure}[thb]
\centering
\includegraphics[width=3.5in]{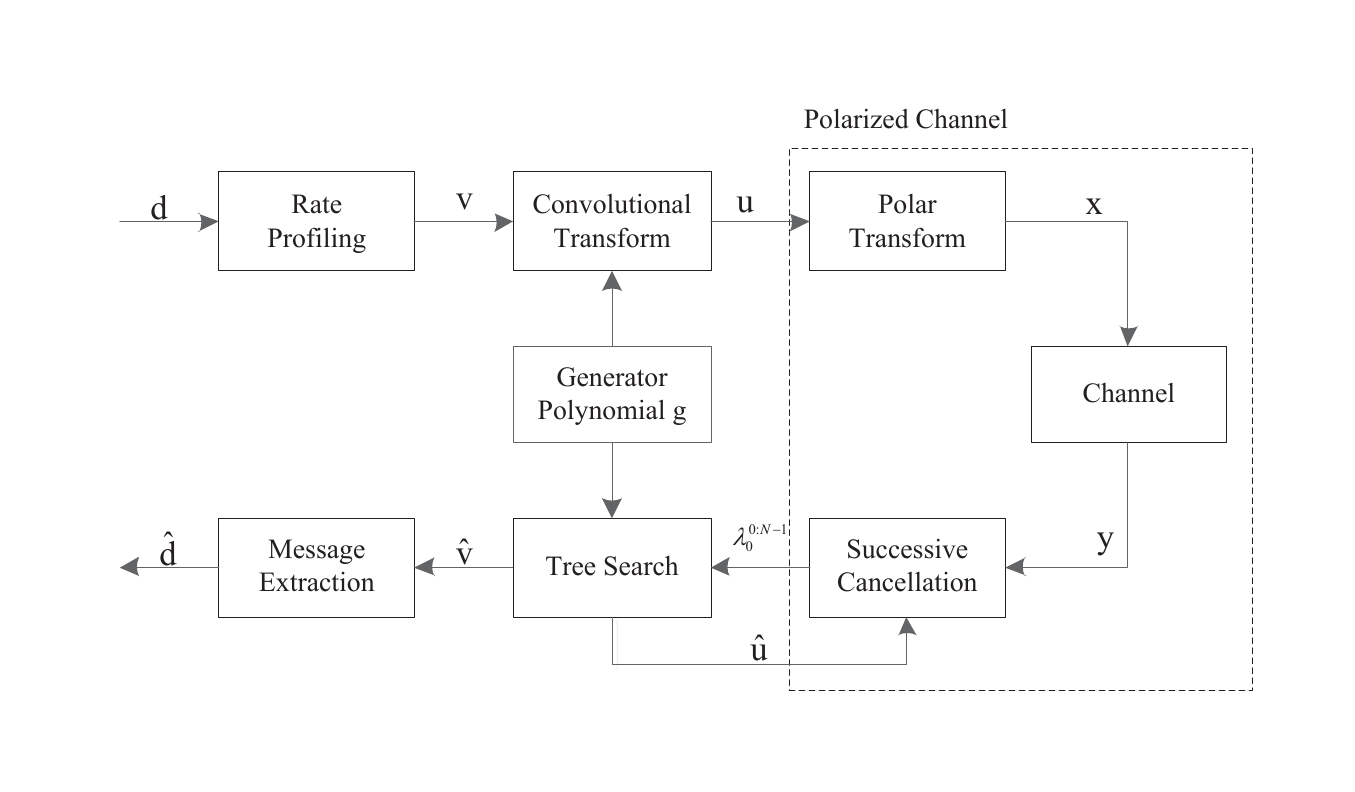}
\caption{Block diagram of PAC coding scheme.}
\label{fig-1}
\end{figure}

Different from conventional convolutional codes, the convolutional transform rate of PAC codes is 1, so it has no error correction capability \cite{ref6}. 
After convolution transform, ${u}_{i},i\in \mathcal{A}^{c}$ is not frozen like polar codes, so that frozen bits can also act as information bits and parity-check bits \cite{ref20}. 
The polar transform in Fig. 1 is the polar encoding, and the codewords $\mathbf{x}={x}^{N-1}_{0}=\{{x}_{0},\cdots,{x}_{N-1}\}$ is 
\begin{equation}
\label{ex10}
{x}^{N-1}_{0}={u}^{N-1}_{0}{G}_{N}.
\end{equation}

Assuming binary phase-shift keying (BPSK) modulation is used to transmit codewords, the vector $\mathbf{y}={y}^{N-1}_{0}=\{{y}_{0},\cdots,{y}_{N-1}\}$ is received after passing through the channel. 
The channel LLRs ${\lambda}^{i}_{n}=\frac{2{y}_{i}}{{\sigma}^{2}}$ can be calculated according to the receive vector $\mathbf{y}$. 
The outputs of the successive cancellation process are denoted by ${\lambda}^{0:N-1}_{0}$. 
The tree search algorithm in Fig. 1 can be list decoding \cite{ref7}, \cite{ref8}, Fano decoding \cite{ref7}, \cite{ref21}, list Viterbi decoding \cite{ref22}, stack decoding, etc. 
The estimated $\mathbf{\hat{v}}={\hat{v}}^{N-1}_{0}=\{{\hat{v}}_{0},\cdots,{\hat{v}}_{N-1}\}$ of $\mathbf{v}$ is obtained by the tree search algorithm. 
Finally, the estimated $\mathbf{\hat{d}}={\hat{d}}^{K-1}_{0}=\{{\hat{d}}_{0},\cdots,{\hat{d}}_{K-1}\}$ of the information vector is extracted in message extraction. 

\subsection{List Decoding}

List decoding is a breadth-first search algorithm. 
In a binary tree, each branch of level $i$ corresponds to the decision ${\hat{v}}_{i}=0$ or ${\hat{v}}_{i}=1$. 
List decoding only needs to trace $L$ paths, and in the $i$ level, the path is decided according to ML rule $h({\lambda}^{i}_{0}[l])$, where $l\in\{1,\cdots,L\}$. 
ML rule is 
\begin{equation}
\label{ex11}
h({\lambda}^{i}_{0}[l]) = \begin{cases}
{0},&{{\lambda}^{i}_{0}[l]>0,} \\ 
{1},&{otherwise.} 
\end{cases}
\end{equation}

If $i\not\in\mathcal{A}$, ${\hat{v}}_{i}[l]=0$, where $l\in\{1,\cdots,L\}$. 
If $i\in\mathcal{A}$, then a path splits into two paths ${\hat{v}}_{i}[l]=0$ and ${\hat{v}}_{i}[{l}^{\prime}]=1$, where $l\in\{1,\cdots,L\}$, ${l}^{\prime}\in\{L+1,\cdots,2L\}$. 
After path splitting, path metric (PM) are used to determine the $L$ paths with the smallest PM value. 
PM value is 
\begin{equation}
\label{ex12}
{PM}^{(i)}_{l} = \begin{cases}
{{PM}^{(i-1)}_{l}+|{\lambda}^{i}_{0}[l]|},&{if\ {\hat{u}}_{i}[l]\neq h({\lambda}^{i}_{0}[l]),} \\ 
{{PM}^{(i-1)}_{l}},&{otherwise.} 
\end{cases}
\end{equation} 
Where, ${\hat{u}}_{i}[l]=\sum^{m}_{j=0} {{g}_{j}}{{\hat{v}}_{i-j}[l]}$. 

Finally, at level $N-1$ , the path with the smallest PM value is selected as the estimation codeword. 

\subsection{Fano Decoding}

Fano decoding explores the most promising paths on a node that can be moved to ancestor nodes or child nodes. 
If the ancestor node is reached, the process is backward traversal; otherwise, it is forward traversal. 
When forward traversal, Fano decoding steps are similar to SC decoding. 
In backward traversal, partial rewind of SC algorithm is used to avoid decoding from scratch \cite{ref7}, \cite{ref21}. 
\begin{figure}[thb]
\centering
\includegraphics[width=3.5in]{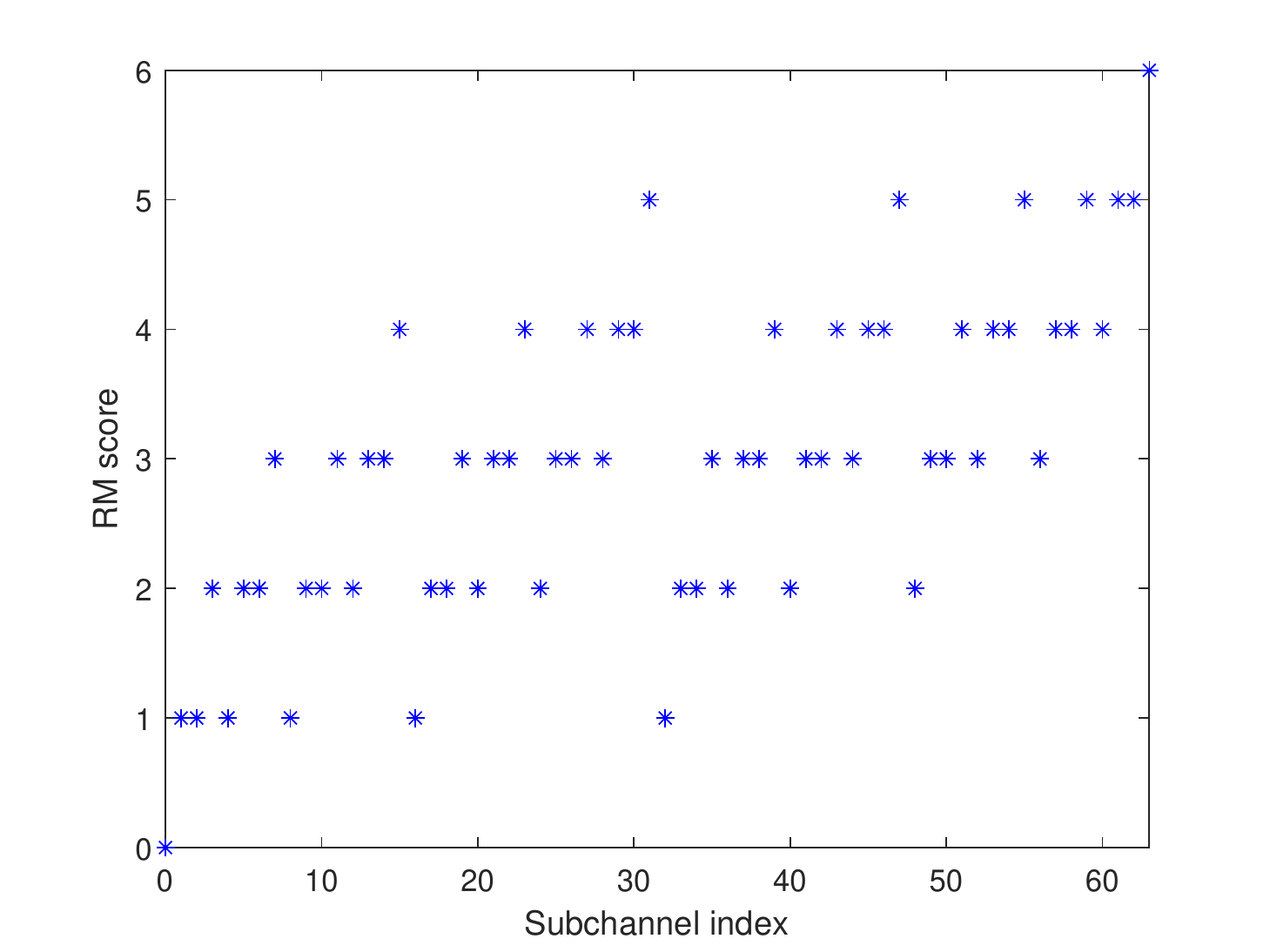}
\caption{Distribution of RM score when $N=64$.}
\label{fig-2}
\end{figure}

Fano decoding has excellent BLER performance, but the average computational complexity is variable. 
Fano decoding uses threshold spacing $\Delta$ to control the performance-complexity tradeoff. 
Choosing a smaller value $\Delta$ will cause Fano decoding to do a lot of backward traversal. 
Although a smaller value $\Delta$ can improve the BLER performance of Fano decoding, its average computational complexity also increases dramatically. 
Fano metric can provide an indication of whether to the current node's partial path be correct, so choosing an appropriate Fano metric can reduce backward traversal and the average computational complexity. 
An optimal metric function on average \cite{ref23} is calculated as 
\begin{equation}
\label{ex13}
{\Gamma}({u}^{i}_{0})={\Gamma}({u}^{i-1}_{0})+1+{\log}_{2}{P({y}^{N-1}_{0},{u}^{i-1}_{0}|{u}_{i})}-{b}_{i}.
\end{equation}
Where, ${P({y}^{N-1}_{0},{u}^{i-1}_{0}|{u}_{i})}$ is the a posterior probability of ${u}_{i}$. 
For better BLER performance and lower computational complexity, bias values ${b}_{i}$ is satisfied 
\begin{equation}
\label{ex14}
0\leqslant{b}_{i}\leqslant I({W}^{(i)}_{N}).
\end{equation} 
bias values ${b}_{i}={E}_{0}(1,{W}^{(i)}_{N})$ \cite{ref23}.

\section{The List-Search Based Construction}
In this section, the LS based construction of PAC codes is proposed. 
Many decoding algorithms of PAC codes can be regarded as ML decoding. 
The bottleneck of ML decoding is the number of minimum weight codewords, and its BLER performance is related to weight spectrum. 
In addition, LS rate-profiles takes into account the limited average computational complexity of sequence decoding for PAC codes, so that the constructed codes can be tradeoff between decoding performance and complexity. 

\subsection{RM rate-profiles}
The most important parameter of RM rate-profiles is RM score function. 
RM rate-profiles are similar to the polarization weight (PW) rate-profiles \cite{ref24}, and the selection of information subchannel is only related to the subchannel index. 
Consider the subchannel index $i$, $0\leqslant i <N$ and the binary representation is $i\triangleq \{{b}_{n-1},\cdots,{b}_{0}\}$. 
RM score $s(i)$ is the number of ones in the binary representation of the subchannel index. 
RM score is 
\begin{equation}
\label{ex15}
s(i)=\sum^{n-1}_{j=0} {{b}_{j}}.
\end{equation} 
For example, if the binary representation of subchannel index 5 is $\{1,0,1\}$, then $s(5)=2$. 
Therefore, the $K$ subchannel with the largest RM score is selected as the information subchannel. 
It can be seen from the RM score formula that RM score is a special case of PW formula. 

RM rate-profiles can only be constructed under specific parameters, and the number of information bits $K$ is 
\begin{equation}
\label{ex16}
K=\sum^{n}_{q=r} {\dbinom{n}{q}}.
\end{equation} 
The range of $r$ is $0\leqslant r \leqslant n$. 
The distribution of RM score is shown in Figure 2. 
There are a large number of subchannels with the distribution of same RM score, and the number of the same RM score is symmetric, that is, the number of RM score $r$ is the same as that of RM score $n-r$. 
\begin{figure}[htb]
\centering
\includegraphics[width=2.3in]{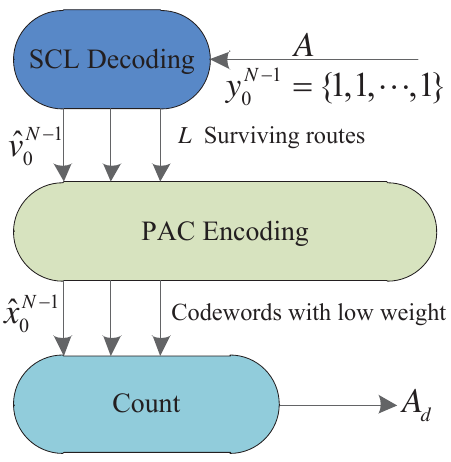}
\caption{SCL decoding to obtain partial weight spectrum of PAC codes block diagram.}
\label{fig-3}
\end{figure}

If $N=64$ and $K=22$, the subchannels with RM score between 4 and 6  are selected as the information subchannels. 
If $N=64$ and $K=32$, RM rate-profiles can not be gotten. 
RM-polar rate-profiles selects the subchannels with RM score between 4 and 6 as the information subchannels, and selects half of the subchannels with RM score 3 as the information sub-channels according to the reliability of each subchannel. 
The BLER performance of RM-polar rate-profiles is not ideal, and its BLER performance deviates seriously from the dispersion bound. 

\subsection{Weight Spectrum} 
For a linear block code with the code length $N$ and the number of information bits $K$, its weight enumeration function (WEF) is 
\begin{equation}
\label{ex17}
A(Z)=\sum^{N}_{d=0} {{A}_{d}{Z}^{d}}.
\end{equation} 
Where, ${A}_{d}$ indicates the number of codewords whose weight is $d$, total number of codewords is ${2}^{K}$. 
\begin{figure}[t]
\centering
\includegraphics[width=3.5in]{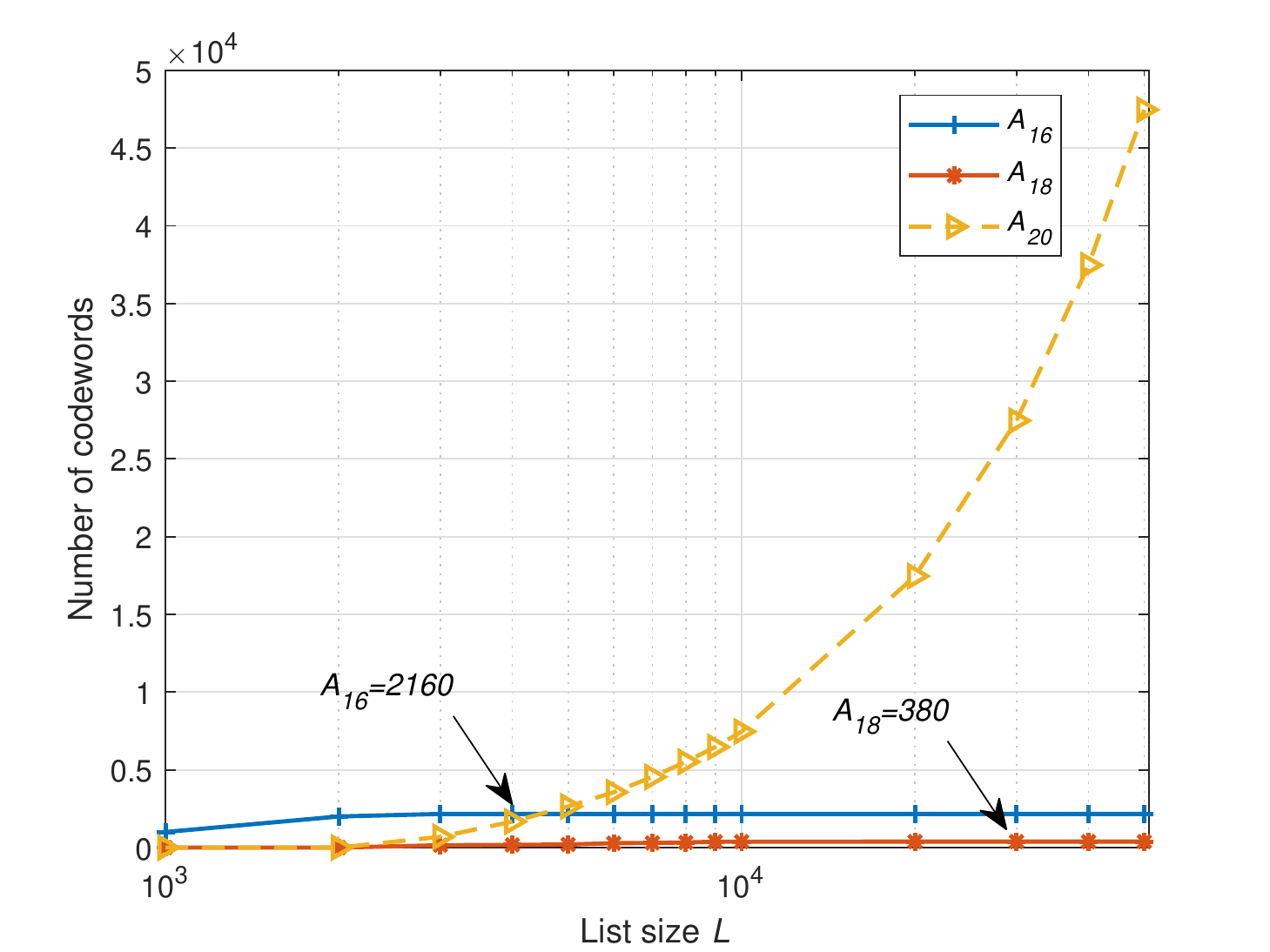}
\caption{Partial weight spectrum of PAC(128,64) codes versus the list size.}
\label{fig-4}
\end{figure}

A lemma of weight spectrum of polar codes is given in \cite{ref25}: If the number of 1 in the binary of index $i$ is represented by $s(i)$ and the row weight of $i$ of the polar generation matrix ${G}_{N}$ is ${2}^{s(i)}$, then the minimum Hamming weight of the polar codes is equal to the minimum row weight of the information bits in ${G}_{N}$. 
As a concatenated code of polar codes, PAC codes combines the rate profiling and convolutional transform in Figure 1 to provide a generalization of rate profiling. 
Thus the minimum Hamming weight of PAC codes is equal to the minimum row weight in the convolutional transform codewords ${u}^{N-1}_{0}$. 

Many decoding algorithms of PAC codes can be regarded as ML decoding, and the bottleneck of ML decoding is the number of minimum weight codewords. 
The BLER performance of PAC codes can be analyzed by partial weight spectrum. 
Li et al. proposed to use SCL decoding to obtain partial weight spectrum of polar codes \cite{ref26}. 
This method can obtain the number of minimum weight codewords. 

With appropriate modifications to SCL decoding and polar encoding in \cite{ref26}, partial weight spectrum of PAC codes can be obtained. 
Figure 3 shows the steps to obtain partial weight spectrum of PAC codes with SCL decoding. 
The steps are as follows 
\begin{enumerate}{}{}
\item{Transmission all zeros codeword over noiseless AWGN channel.}. 
\item{The estimation ${\hat{v}}^{N-1}_{0}$ of $L$ surviving paths of source word ${v}^{N-1}_{0}$ is obtained using SCL decoding.}
\item{PAC encoding (convolution transform and polar transform) is performed on the estimated ${\hat{v}}^{N-1}_{0}$ of $L$ source words to obtain $L$ codewords ${\hat{x}}^{N-1}_{0}$.}
\item{Count the weight and the number of weight for $L$ codewords.}
\end{enumerate}

The proof to obtain partial weight spectrum of polar codes using SCL decoding is given in \cite{ref27}. 
Similarly, this method can be applicable to PAC codes. 
When all zeros codeword passes through a noiseless AWGN channel using BPSK modulation, each path probability among the decoding tree is inversely proportional to the weight of the corresponding codeword. 
This indicates that the codewords with smaller weight are more probable to be kept in the candidated codeword list. 
\begin{figure}[t]
\centering
\includegraphics[width=3.5in]{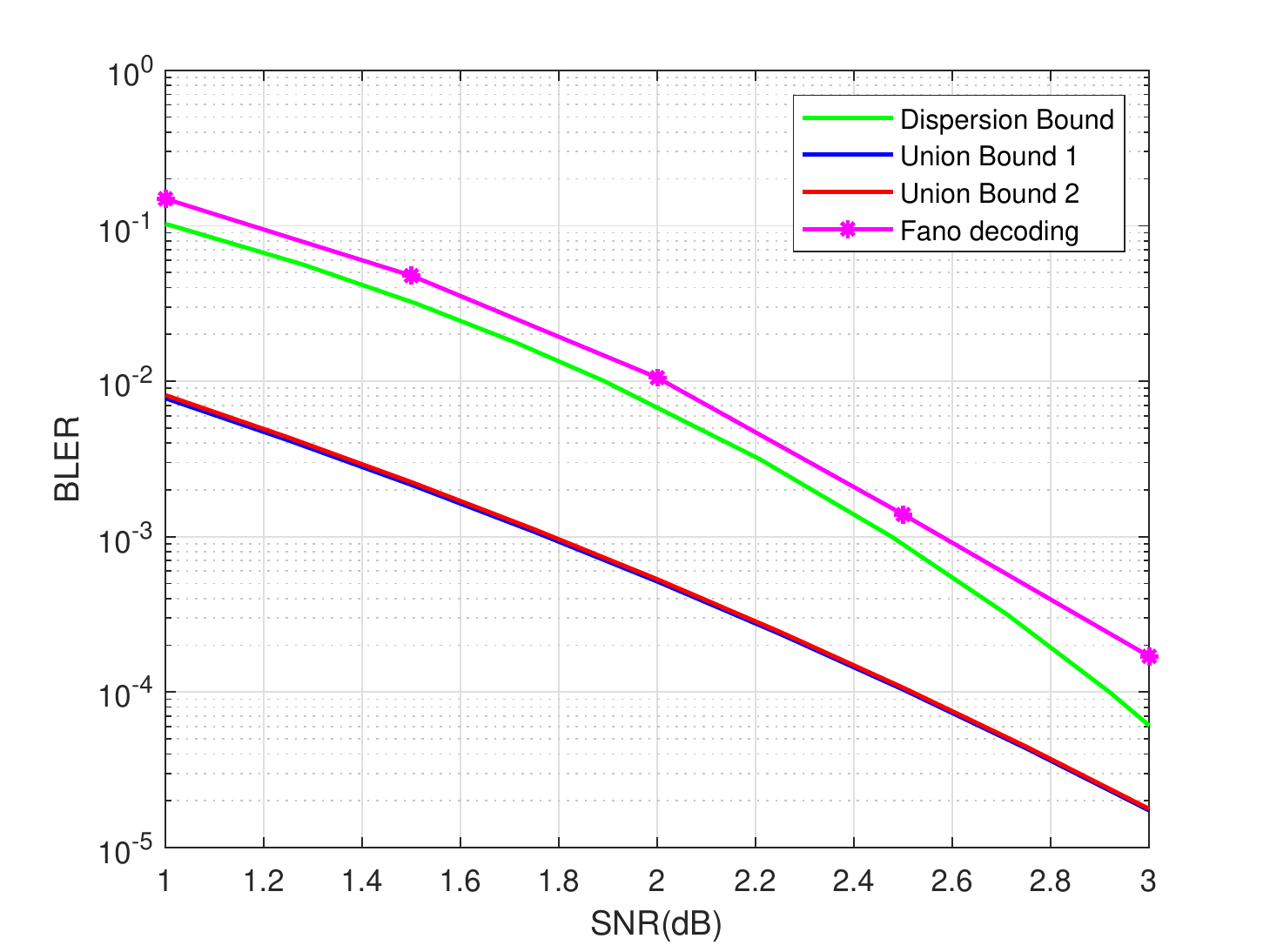}
\caption{BLER performance of Fano decoding for PAC(128,64) codes.}
\label{fig-5}
\end{figure}

Figure 4 shows partial weight spectrum of PAC(128,64) codes versus the list size with RM rate-profiles and convolution polynomial $\mathbf{g}=\{1,1,0,1,0,0,0,1,0,0,1\}$(3211 in octal notation). 
It can be seen from the figure 4 that the minimum weight is ${d}_{min}=16$, and the three weights are 16, 18 and 20. 
The number of the codewords with weight 16 is 2160, which remains almost unchanged if list size is greater than 4000. 
Similarly, the number of the codewords with weight 18 is 380, which remains almost unchanged if list size is greater than 30000. 
Therefore, the appropriate list size can be selected to obtain the corresponding weight and the number of weight. 

Given the weight spectrum (or partial weight spectrum) of PAC codes, the BLER performance of ML decoding in the AWGN channel can be computed from the Union Bound ${P}_{UB}$ 
\begin{equation}
\label{ex18}
BLER\leqslant {P}_{UB}=\sum^{N}_{d={d}_{min}} {{A}_{d}{Q(\sqrt{2dR{{E}_{b}/{N}_{0}}})}}.
\end{equation} 
Where, ${d}_{min}$ is the minimum weight, ${A}_{d}$ is the number of codewords with a weight $d$, $R$ is the rate, and ${E}_{b}/{N}_{0}$ is the SNR. 

In the high SNR region, the union bound ${P}_{UB}$ can be approximated as 
\begin{equation}
\label{ex19}
BLER\approx {P}_{UB}={{A}_{{d}_{min}}{Q(\sqrt{2{d}_{min}R{{E}_{b}/{N}_{0}}})}}.
\end{equation} 
Therefore, when the minimum weight is larger and the number of the codewords with minimum weight is smaller, the BLER performance of ML decoding becomes better. 

According to (18) and (19), the union bound is the upper bound of BLER performance of ML decoding , so the BLER performance of ML decoding is better than the union bound. 
As shown in Figure 4, ${A}_{16}=2160$ and ${A}_{18}=380$. 
The union bound 1 in Figure 5 is obtained by the weight 16, and the union bound 2 is obtained by the weights 16 and 18. 
The dispersion bound is an estimate of the average ML decoding performance over the BI-AWGN channel of a code chosen uniformly at random from the ensemble of all possible binary codes. 
The BLER performance of Fano decoding is close to the dispersion bound and above union bound, because the union bound in the figure 5 is approximated (26). 
As the weight increases, the union bound gradually rises above the BLER performance curve. 
\begin{figure}[htb]
\centering
\includegraphics[width=3.5in]{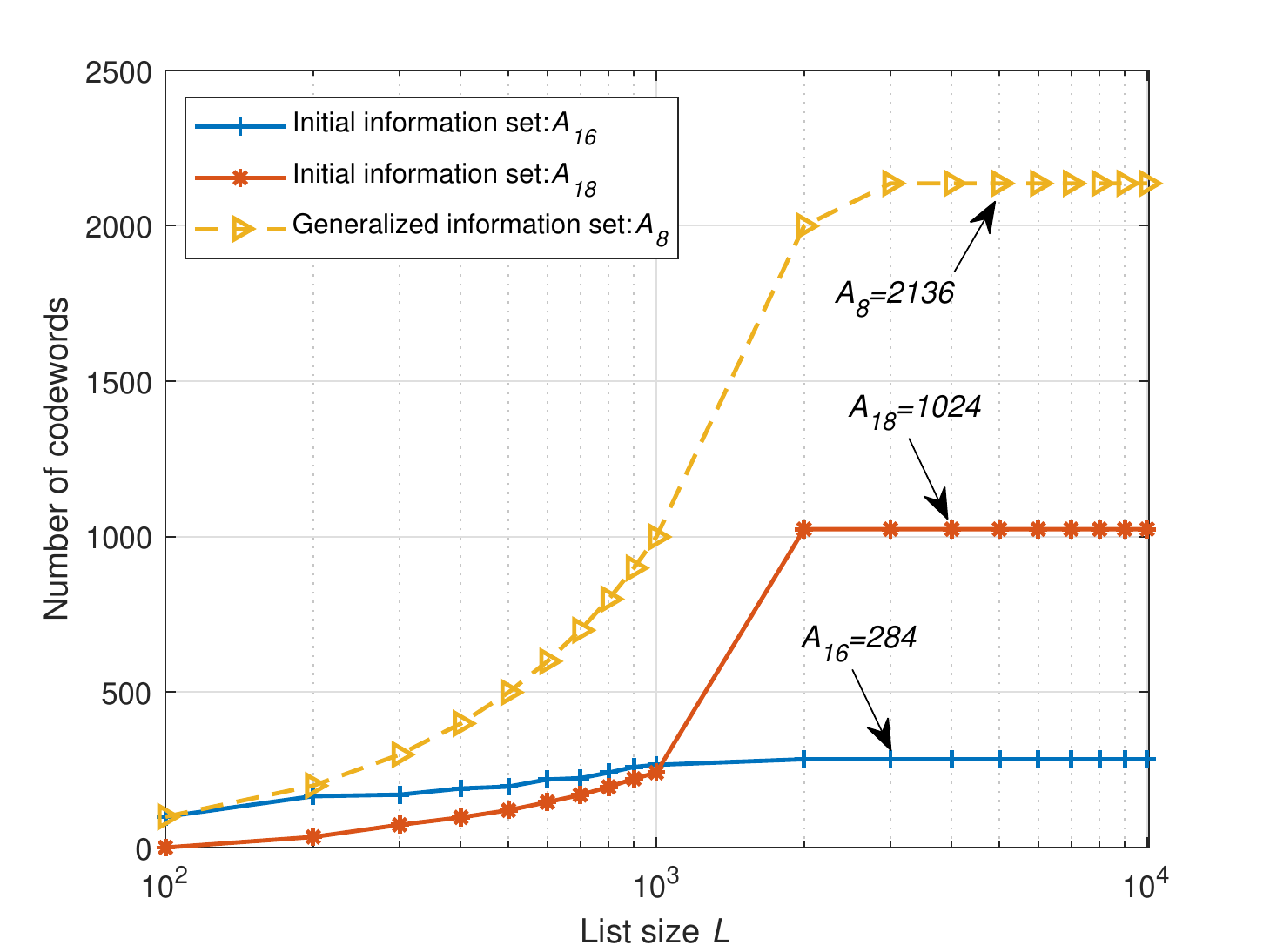}
\caption{Partial weight spectrums of initial information set and generalized information set versus the list size for PAC(64,32) codes.}
\label{fig-6}
\end{figure}

\subsection{Polarized profile}
If the rate $R(W)$ is greater than the cutoff rate ${E}_{0}(1,W)$, $R(W)>{E}_{0}(1,W)$, the complexity of sequence decoding is unacceptable in practice. 

The polarized capacity profile of index $i$ is defined as the sequence $I({W}^{(i)}_{N})$ of cumulatives 
\begin{equation}
\begin{matrix}
\label{ex20}
\sum^{i}_{j=0} {I({W}^{(i)}_{N})}
\end{matrix}
.
\end{equation} 

Similarly, the polarized cutoff rate profile and the polarized rate profile of index $i$ are respectively 
\begin{equation}
\begin{matrix}
\label{ex21}
\sum^{i}_{j=0} {{E}_{0}({W}^{(i)}_{N})}
\end{matrix}
.
\end{equation} 
\begin{equation}
\begin{matrix}
\label{ex22}
\sum^{i}_{j=0} {R({W}^{(i)}_{N})}
\end{matrix}
.
\end{equation} 

The condition of sequence decoding for PAC codes with finite average computational complexity\cite{ref12} is 
\begin{equation}
\begin{matrix}
\label{ex23}
\sum^{i}_{j=0} {R({W}^{(i)}_{N})} < \sum^{i}_{j=0} {{E}_{0}({W}^{(i)}_{N})}
\end{matrix}
.
\end{equation} 
If the polar cutoff rate is obtained when the SNR value is equal to ${E}_{b}/{N}_{0}$, the rate-profiles satisfying (23) has low average computational complexity for SNR values beyond ${E}_{b}/{N}_{0}$. 

\subsection{The construction principle with List-Search}
Given $N$ and $K$, $n={\log}_{2}N$, there is 
\begin{equation}
\label{ex24}
\sum^{n}_{q=r+1} {\dbinom{n}{q}} \leqslant K < \sum^{n}_{q=r} {\dbinom{n}{q}}
\end{equation} 
Where, the range of $r$ is $0 \leqslant r < n$. 

\begin{figure}[htb]
\centering
\includegraphics[width=3.5in]{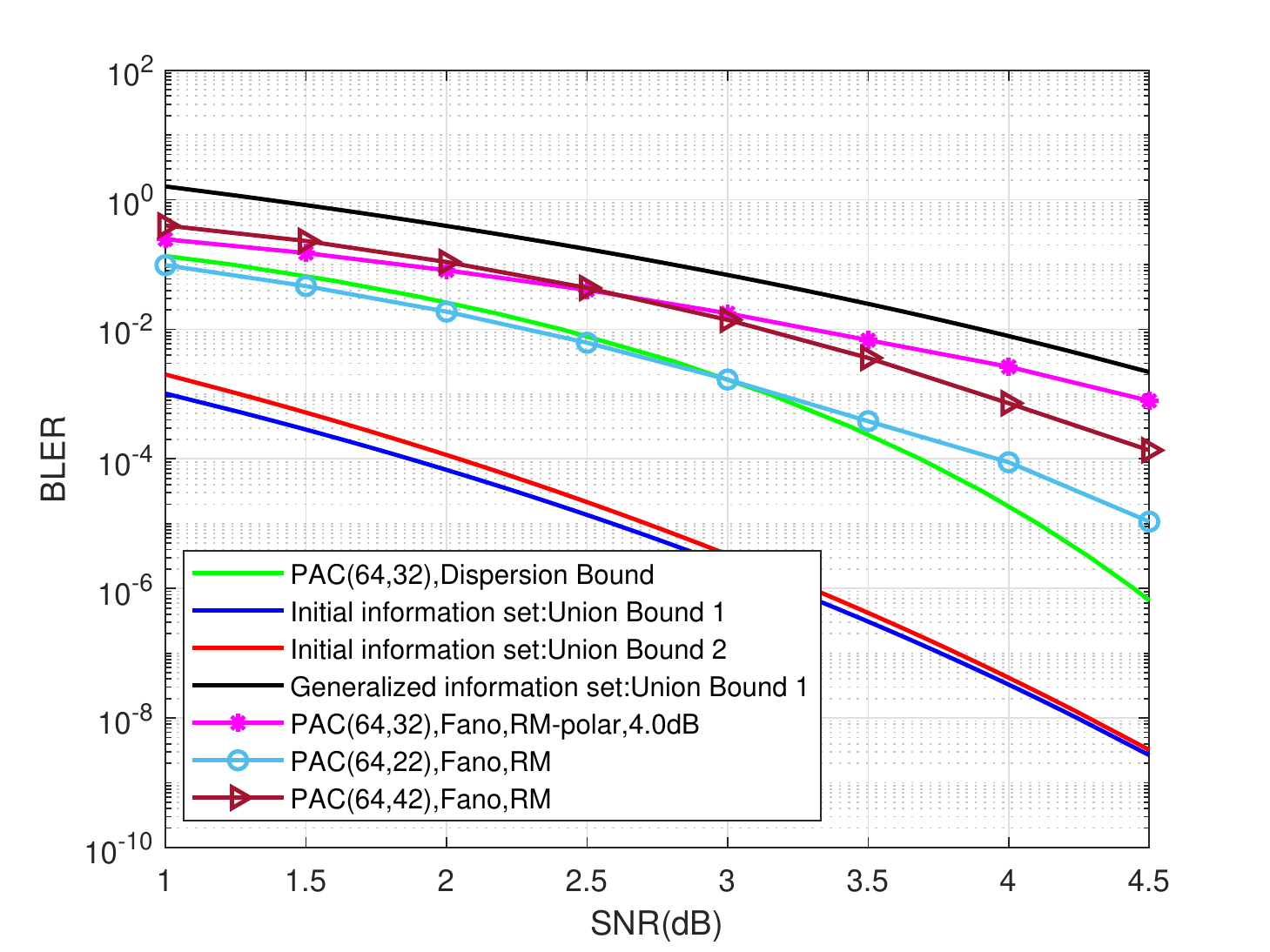}
\caption{BLER performance of Fano decoding for PAC codes.}
\label{fig-7}
\end{figure}

\begin{algorithm}[t]
\caption{The List-Search Based Construction method}
\label{algorithm:alpha_1}
\SetKwData{In}{\textbf{in}}\SetKwData{To}{to}
\DontPrintSemicolon
\SetAlgoLined
\KwIn {$N$, $K$, $\mathbf{g}$, $L$, $Lg$, ${E}_{b}/{N}_{0}$}
\KwOut {$\mathcal{A}$}
$(\mathcal{A},\mathcal{B},\mathcal{C})=RMScore(N,K)$ \;
\If {$|\mathcal{A}| = K$}{
      \Return{$\mathcal{A}$} \;
    }
$({\hat{v}}^{N-1}_{0}[{l}^{\prime}],{\hat{x}}^{N-1}_{0}[{l}^{\prime}],{z}[{l}^{\prime}])=SCLGetWS(\mathbf{g},\mathcal{C},L)$ \;
$(\mathcal{A}[l],\mathcal{B}[l],{\mathbf{M}}^{N}_{0}[l], Lm)=RootInit(\mathcal{A},\mathcal{B},{\hat{v}}^{N-1}_{0}[{l}^{\prime}],{z}[{l}^{\prime}])$ \;
$i \gets 0$ \;
\While{$i<(K-|\mathcal{A}|-1)$}{
  $\mathcal{A}0[{l}^{\prime \prime}] \gets \varnothing, \mathcal{B}0[{l}^{\prime \prime}] \gets \varnothing, {\mathbf{M}0}^{N}_{0}[{l}^{\prime \prime}]  \gets \varnothing$ \;
   \For{$j \gets 0$ \To $Lm-1$}{
    $(\mathcal{A}[j][{l}^{\prime \prime \prime}], \mathcal{B}[j][{l}^{\prime \prime \prime}], {\mathbf{M}}^{N}_{0}[j][{l}^{\prime \prime \prime}])=Split(\mathcal{A}[l],\mathcal{B}[l],{\mathbf{M}}^{N}_{0}[l])$ \;
    $\mathcal{A}0[{l}^{\prime \prime}]=\mathcal{A}0[{l}^{\prime \prime}] \cup \mathcal{A}[j][{l}^{\prime \prime \prime}]$ \;
    $\mathcal{B}0[{l}^{\prime \prime}]=\mathcal{B}0[{l}^{\prime \prime}] \cup \mathcal{B}[j][{l}^{\prime \prime \prime}]$ \;
    ${\mathbf{M}0}^{N}_{0}[{l}^{\prime \prime}]={\mathbf{M}0}^{N}_{0}[{l}^{\prime \prime}] \cup {\mathbf{M}}^{N}_{0}[j][{l}^{\prime \prime \prime}]$ \;
  }
  $(\mathcal{A}[l],\mathcal{B}[l],{\mathbf{M}}^{N}_{0}[l], Lm)=Pruning(\mathcal{A}0[{l}^{\prime \prime}], \mathcal{A}0[{l}^{\prime \prime}], \mathcal{B}0[{l}^{\prime \prime}], Lg)$ \;
$i \gets i+1$ \;
}
\Return{$\mathcal{A}$}
\end{algorithm}

The initial information set $\mathcal{A}$ and the optional information set $\mathcal{B}$ can be determined by the RM score. 
The initial information set $\mathcal{A}$ contains all indexes of RM scores $s(i)>r$, and the optional information set $\mathcal{B}$ contains all indexes of RM scores $s(i)=r$. 
The Generalized information set $\mathcal{C}=\mathcal{A}\cup\mathcal{B}$. 
As shown in Figure 2, $|\mathcal{A}|=22$, $|\mathcal{B}|=20$, $|\mathcal{C}|=42$. 
It can be seen that the minimum weight of PAC(64,42) codes is 8, and that of PAC(64,22) codes is 16 in Figure 6. 
For the information set of PAC(64,32) codes with LS construction, 10 indexes is selected from the set $\mathcal{B}$ and merged to the set $\mathcal{A}$, so the minimum weight of PAC(64,32) codes is between 8 and 16. 
Therefore, if the minimum weight is as big as possible, BLER performance of PAC(64,32) codes will be improved. 
The union bound 1 in Figure 7 is caculated with the minimum weight, and the union bound 2 is caculated with the minimum weight and the second least weight. 
At a high SNR, the BLER performance of Fano decoding for PAC(64,42) codes using RM rate-profiles in Figure 7 is better than that of Fano decoding for PAC(64,32) codes using RM-polar rate-profiles. 
We can conclude that the BLER performance of the constructed PAC codes with RM-polar is not optimal. 

If the rates coincide, the LS rate-profiles and the RM rate-profiles will yield the same codes. 
For LS Based construction of PAC($N$,$K$) codes, assuming $Lg$ is the list size of LS, the $l$-th LS metric 
\begin{equation}
\label{ex25}
{\mathbf{M}}^{N}_{0}[l]=\{{\mathbf{M}}_{0}[l],\cdots,{\mathbf{M}}_{i}[l],\cdots,{\mathbf{M}}_{N}[l]\}
\end{equation} 
Where, $0\leqslant l<Lg$, ${\mathbf{M}}^{N}_{0}[l]$ is the statistical partial weight spectrum of the $l$-th list by LS construction, and ${\mathbf{M}}_{i}[l]$ is the number of the counted codewords when the weight is $i$. Let $0\leqslant k \leqslant N$, if each element ${\mathbf{M}}_{k}[i]$ in ${\mathbf{M}}^{N}_{0}[i]$ is equal to ${\mathbf{M}}_{k}[j]$ in ${\mathbf{M}}^{N}_{0}[j]$, ${\mathbf{M}}^{N}_{0}[i]$ is equal to ${\mathbf{M}}^{N}_{0}[j]$. 
If ${\mathbf{M}}_{k}[i]$ is larger than ${\mathbf{M}}_{k}[j]$, and ${\mathbf{M}}^{k-1}_{0}[i]$ is equal to ${\mathbf{M}}^{k-1}_{0}[j]$, ${\mathbf{M}}^{N}_{0}[i]$ is larger than ${\mathbf{M}}^{N}_{0}[j]$. 
For example, LS metric $\{0,5,2,0\}$ and LS metric $\{0,5,2,1\}$, where LS metric $\{0,5,2,1\}$ are larger than LS metric $\{0,5,2,0\}$. 

\begin{table*}[t]
\caption{Rate profiles of the LS based construction
\label{tab:table1}}
\centering
\begin{tabular}{c c c c c c}
\hline
$(N,K)$ & $\mathbf{g}$ & $L$ & $Lg$ & ${E}_{b}/{N}_{0}$(dB) & Rate Profile($\mathbf{\alpha}$) \\
\hline
\multirow{6}{*}{$(256,128)$} & \multirow{3}{*}{$\mathbf{g1}=0o3211$} & 40000 & 400 & 2.5 & 00000001000305770013077F1757577F0013075F17773FFF175F177F177F7FFF \\
\cline{3-6}
 &  & 40000 & 400 & 3.0 & 00000003000317570013075F1757577F0013075F17773F7F175F177F177F7FFF \\
\cline{3-6}
 &  & 40000 & 400 & 3.2 & 00000005001317170013075F1757577F0013075F17773F7F175F177F177F7FFF \\
\cline{2-6}
 & \multirow{2}{*}{$\mathbf{g2}=0o133$} & 40000 & 400 & 2.5 & 000000010005033F0015155717577FFF0015155707777FFF171737FF177F7FFF \\
\cline{3-6}
 &  & 40000 & 400 & 3.2 & 00000005001511770015155717577F7F00151557133F7FFF1717377F177F7FFF \\
\cline{2-6}
 & \multirow{1}{*}{$\mathbf{g3}=0o1$} & 200000 & 400 & 3.2 & 000000050011031700031117051717FF01031557155F7FFF177F7FFF7FFFFFFF \\
\hline
\multirow{3}{*}{$(128,42)$} & \multirow{3}{*}{$\mathbf{g1}=0o3211$} & 40000 & 400 & 2.5 & 0000000300130757001307171717177F \\
\cline{3-6}
 &  & 40000 & 400 & 3.0 & 0000000500130757001307171717177F \\ 
\cline{3-6}
 &  & 40000 & 400 & 3.5 & 0000001500130357001307171717177F \\
\hline
\multirow{3}{*}{$(128,85)$} & \multirow{3}{*}{$\mathbf{g1}=0o3211$} & 40000 & 400 & 2.5 & 0001133F077F7FFF173F7F7F177F7FFF \\
\cline{3-6}
 &  & 40000 & 400 & 3.0 & 0003077F177F7FFF171F377F177F7FFF \\ 
\cline{3-6}
 &  & 40000 & 400 & 3.5 & 0013077F177F3FFF175F177F177F7FFF \\
\hline
\multirow{2}{*}{$(64,32)$} & \multirow{2}{*}{$\mathbf{g1}=0o3211$} & 40000 & 400 & 2.5 & 0003157F171F177F \\
\cline{3-6}
 &  & 40000 & 400 & 3.0 & 0007177F1517177F \\
\cline{2-6}
 & \multirow{2}{*}{$\mathbf{g2}=0o133$} & 40000 & 400 & 2.5 & 0005077F1337577F \\
\cline{3-6}
 &  & 40000 & 400 & 3.0 & 0013077F1337177F \\
\cline{2-6}
 & \multirow{1}{*}{$\mathbf{g3}=0o1$} & 100000 & 400 & 2.5 & 00051357153F1FFF \\
\hline
\end{tabular}
\end{table*}

The LS based construction is detailed in Algorithm 1, where $N$ is the code length, $K$ is the number of information bits, $\mathbf{g}$ is convolution polynomial, $L$ is the SCL decoding list size, $Lg$ is the LS list size, and ${E}_{b}/{N}_{0}$ is the SNR. 
The $RootScore()$ function in line 1 determines the initial information set $\mathcal{A}$ and the optional information set $\mathcal{B}$ from (24), and $\mathcal{C}=\mathcal{A} \cup \mathcal{B}$. 
The $SCLGetWS()$ function in line 5 uses SCL decoding to decode the partial weight spectrum of $\mathcal{C}$, and obtains $L$ source word estimates ${\hat{v}}^{N-1}_{0}[{l}^{\prime}]$ and $L$ codeword estimates ${\hat{x}}^{N-1}_{0}[{l}^{\prime}]$. 
${z}[{l}^{\prime}]$ is  one of the weight of to ${\hat{x}}^{N-1}_{0}[{l}^{\prime}]$, where $0 \leqslant{l}^{\prime} < L$. 
The $RootInit()$ function in line 6 makes $\mathcal{A}[l]=\mathcal{A}[l] \cup \mathcal{B}_{l}$, and $\mathcal{A}[l] \cup \mathcal{B}[l]=\mathcal{C}$, and updates the LS metric ${\mathbf{M}}^{N}_{0}[l]$. 
During LS metric ${\mathbf{M}}^{N}_{0}[l]$ updating, if bit ${\hat{v}}_{{\mathcal{B}}_{l}}[{l}^{\prime}]$ whose index is ${\mathcal{B}}_{l}$ in ${\hat{v}}^{N-1}_{0}[{l}^{\prime}]$ is 1, and other indexes of ${\hat{v}}^{N-1}_{0}[{l}^{\prime}]$ are 1 in $\mathcal{A}[l]$, then the $z[{l}^{\prime}]$-th element ${\mathbf{M}}_{z[{l}^{\prime}]}[l]$ in ${\mathbf{M}}^{N}_{0}[l]$ is added by 1. 
$RootInit()$ function initializes the root node of the LS tree, and root node size is $|{\mathcal{B}}|$. 
Let $Lm=|{\mathcal{B}}|$, $Lm$ is instantaneous list size for LS. 
The $Split()$ function splits the $j$-th parent node in the LS tree into the $|\mathcal{A}[j]|$ child node, and updates the information set of list, optional information set of list and the LS metrics of list. 
The $Pruning()$ function in line 16 keeps only one set of the same information in the list and removes child nodes that do not satisfy (23). 
After the above steps, the $Lm$ child nodes are remained. 
If $Lm>Lg$, the smallest $Lg$ child nodes with LS metric are kept, and the child nodes are sorted by LS metric from smallest to largest. 
Finally, the information set of the smallest LS metric is returned as the information set of the LS construction. 
During path pruning in Line 16, removing duplicate child nodes can make more child nodes to be searched. 
In addition, removing the child nodes that do not satisfy (23) can get the constructed codes with a limited average computational complexity during sequential decoding.

Table I shows rate-profiles of the LS based construction, where the convolution polynomial $\mathbf{g}$ is $\mathbf{g1}=\{1,1,0,1,0,0,0,1,0,0,1\}$ (3211 in octal notation), $\mathbf{g2}=\{1,0,1,1,0,1,1\}$ (133 in octal notation), and $\mathbf{g3}=\{1\}$ (1 in octal notation). 
For compact representation of the rate profiles, we use the binary vector $\mathbf{\alpha}$ of length $N$ such that $\mathbf{\alpha}_{i}=1$ if $i \in \mathcal{A}$ and $\mathbf{\alpha}_{i}=0$, for $0\leqslant i < N$ \cite{ref12}. 
Table I lists the rate-profiles when $(256,128)$, $(128,42)$, $(128,85)$ and $(64,32)$. 
The BLER performance of RM-polar rate-profiles, MC rate-profiles and WS rate-profiles is compared Section V.

\section{The Path-Splitting Critical Sets Based Construction}
\subsection{The Complete Path-Splitting Critical Sets}
The RM rate-profiles and the LS rate-profiles (a generalization of the RM rate-profiles) consider the RM score and the number of minimum weight codewords, so that the BLER performance of the constructed PAC codes with RM or LS rate-profiles is close to the dispersion bound. 
With SCL decoding, PAC codes with the small block lengths can obtain BLER performance close to the dispersion bound \cite{ref7}, \cite{ref8}, \cite{ref12}. 
Pruned SCL decoding \cite{ref16} and PSCL decoding \cite{ref18} can greatly reduce number of sorting operations, but number of sorting operations is variable with SNR value. 
Path splitting (one path divided into two paths) occur on information bits during SCL decoding, Pruned SCL decoding, or PSCL decoding. 
When the number of paths is greater than the list size $L$, the paths are sorted and $L$ paths with the smallest path metric value are kept. 
If the path-splitting critical sets $CS\subset\mathcal{A}$, and the path splitting is performed only on each element of the path-splitting critical sets, then number of sorting operations will be reduced. 
For the information bit $i\in CS$, the decoding path expands along the most reliable child node. 
Reducing the number of sort operations can reduce the delay, so PSCS need to be obtained with a feasible scheme. 
PSCS are different from the flip critical sets in \cite{ref28}, and only path splitting is performed on each element in PSCS during SCL-type decoding. 

\begin{algorithm}[t]
\caption{The Complete Path-Splitting Critical Sets Based Construction method}
\label{algorithm:alpha_2}
\SetKwData{In}{\textbf{in}}\SetKwData{To}{to}
\DontPrintSemicolon
\SetAlgoLined
\KwIn {$N$, $K$, $\mathcal{A}$}
\KwOut {$CPSCS$}
\If{$\begin{matrix} K=\sum^{n}_{q=r} {\dbinom{n}{q}} \end{matrix}$}{
    $CPSCS$ is the set of all indexes in $\mathcal{A}$ where $s(i)=r$ \;
    \Return{$CPSCS$} \;
  }
  $CPSCS$ is the set of all indexes in $\mathcal{A}$ where $s(i)=r$ and $s(i)=r+1$ \;
\Return{$CPSCS$}
\end{algorithm}

\begin{figure}[htb]
\centering
\includegraphics[width=3.5in]{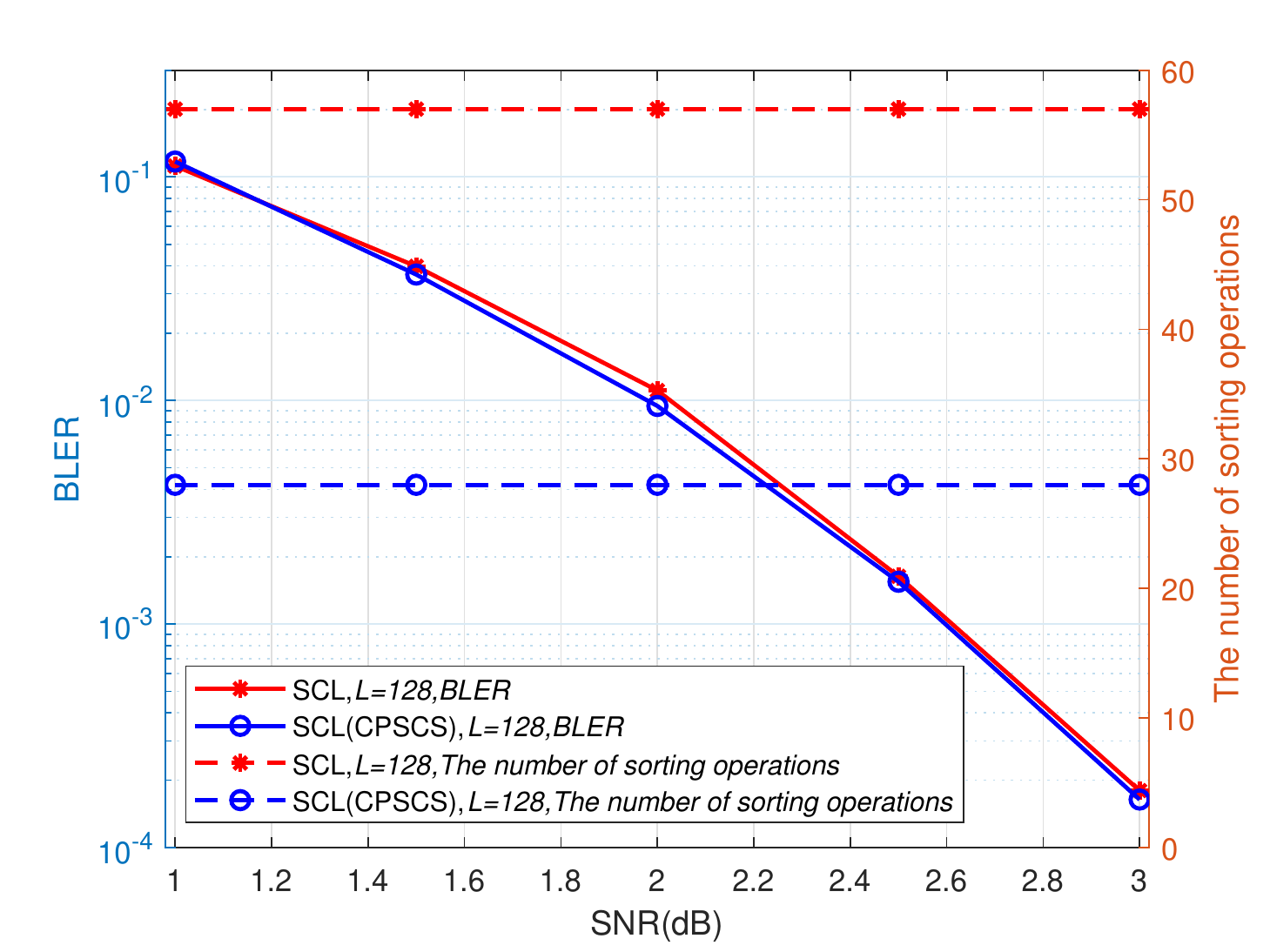}
\caption{BLER performance and the number of sorting operations with SCL decoding and SCL(CPSCS) decoding for PAC(128,64) codes.}
\label{fig-8}
\end{figure}

For an appropriate list size, the BLER performance of SCL decoding approaches the ML bound \cite{ref26}. 
The performance of ML decoding is closely related to the weight spectrum. 
This paper proposes the CPSCS based construction of PAC codes for RM rate-profiles and generalization of RM rate-profiles. 
This method is used to get the CPSCS according to RM score of each information bit. 
The CPSCS based construction of PAC codes is described in Algorithm 2, where $N$ is the code length, $K$ is the number of information bits, $\mathcal{A}$ is the information sets, and $CPSCS$ is CPSCS. 
If $\begin{matrix} K=\sum^{n}_{q=r} {\dbinom{n}{q}} \end{matrix}$, CPSCS is all indexes sets of $s(i)=r$ in $\mathcal{A}$; Otherwise, CPSCS are all indexes sets of $s(i)=r$ and $s(i)=r+1$ in $\mathcal{A}$. 
After SCL(CPSCS) decoding in Figure 8, there is negligible loss in BLER performance, and the number of sort operations is reduced from 57 to 28. 
In Figure 9, for the PAC(64,32) codes of LS-2.5dB rate-profiles, there is negligible loss in BLER performance of SCL(CPSCS) decoding, but the number of sort operations is reduced from 27 to 20. 
The simulation results in Figures 8 and 9 show that CPSCS can reduce the number of sort operations. 

\begin{figure}[t]
\centering
\includegraphics[width=3.5in]{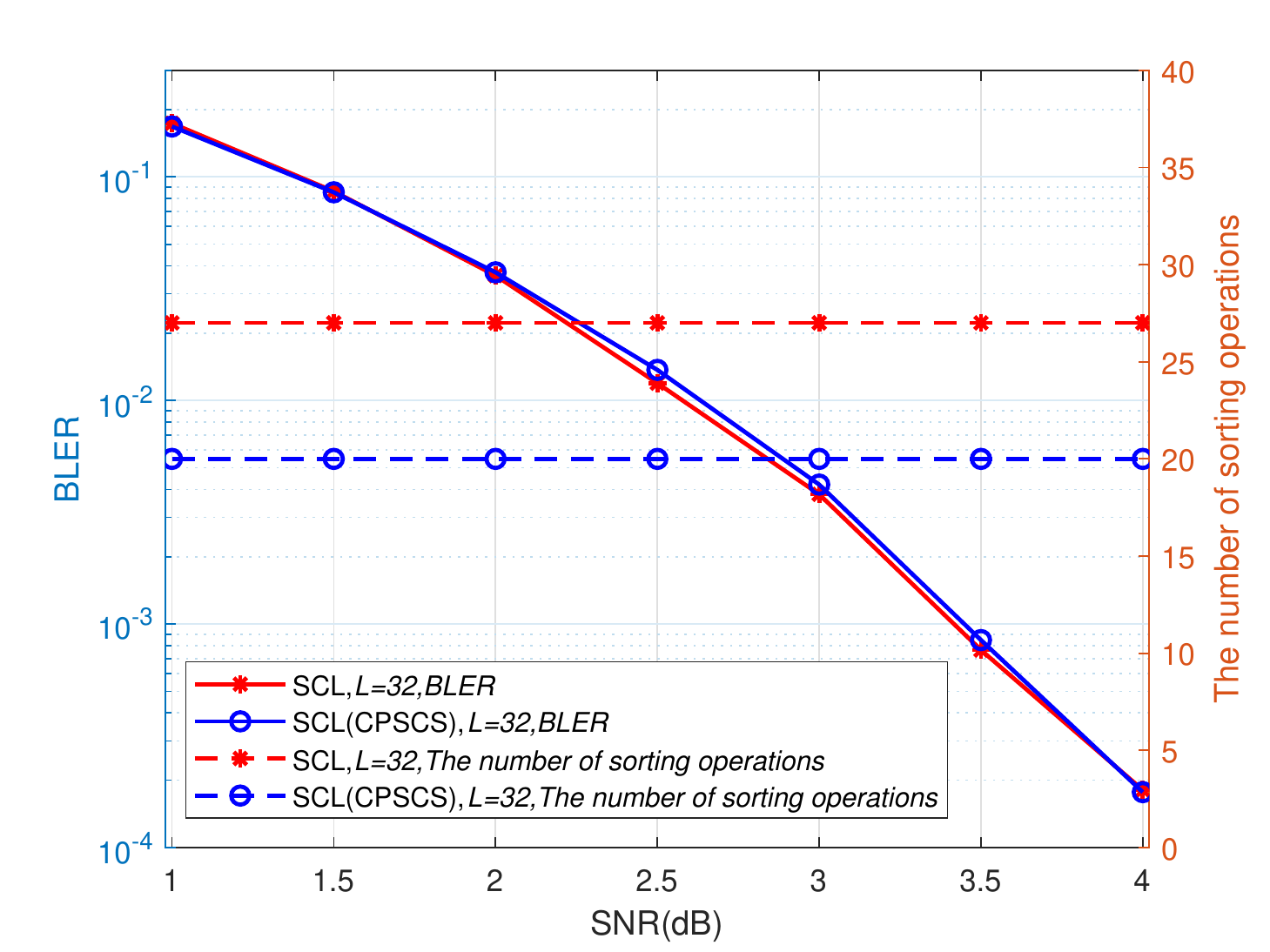}
\caption{BLER performance and the number of sorting operations with SCL decoding and SCL(CPSCS) decoding for PAC(64,32) codes.}
\label{fig-9}
\end{figure}

\subsection{The Path-Splitting Critical Sets}

\begin{figure}[htb]
\centering
\includegraphics[width=3.5in]{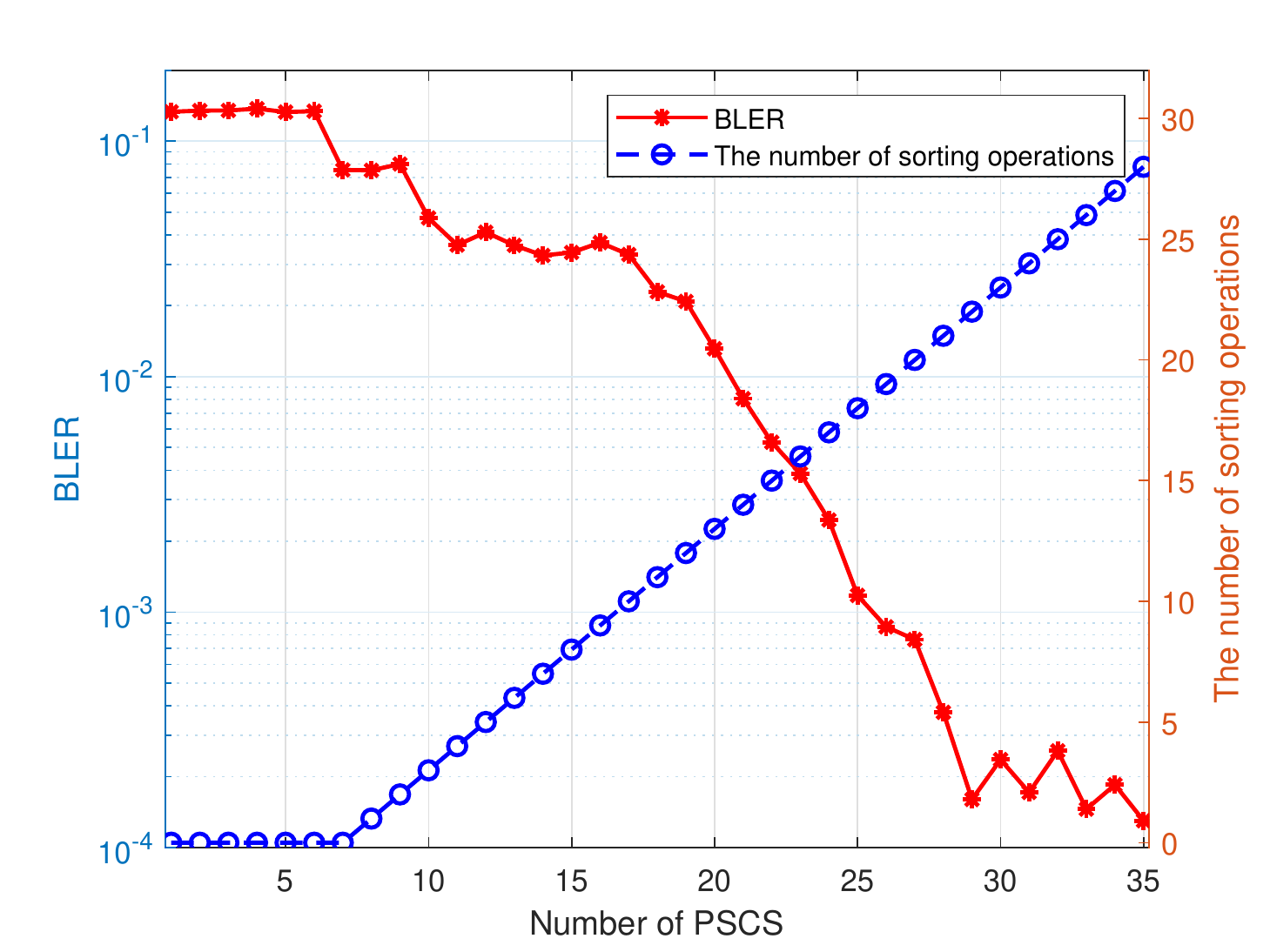}
\caption{BLER performance and the number of sorting operations for SCL decoding at 3dB for PAC(128,64) codes with different number of PSCS.}
\label{fig-10}
\end{figure}

In Figure 9, although CPSCS can reduce the number of sort operations from 27 to 20, this reduction is limited. 
This paper proposes the PSCS based construction to obtain the PSCS from CPSCS according to partial weight spectrum. 
This method can get number of information bits in PSCS, from 1 to $|CPSCS|$. 

\begin{algorithm}[htb!]
\caption{The Path-Splitting Critical Sets Based Construction method}
\label{algorithm:alpha_3}
\SetKwData{In}{\textbf{in}}\SetKwData{To}{to}
\DontPrintSemicolon
\SetAlgoLined
\KwIn {$N$, $K$, $\mathbf{g}$, $\mathcal{A}$, $L$, $Lc$}
\KwOut {$PSCS[l]$}
$({\hat{v}}^{N-1}_{0}[{l}^{\prime}],{\hat{x}}^{N-1}_{0}[{l}^{\prime}],{z}[{l}^{\prime}])=SCLGetWS(\mathbf{g},\mathcal{A},L)$ \;
$(CS1,CS2)=getCS(K,\mathcal{A})$ \;
$(CS[l^{\prime \prime}],{\mathbf{M}}^{N}_{0}[l^{\prime \prime}],Ls)=CSRootInit(CS1,{\hat{v}}^{N-1}_{0}[{l}^{\prime}],{z}[{l}^{\prime}])$ \;
$PSCS[0]=CS[0]$ \;
\For{$i \gets 1$ \To $|CS1|-1$}{
    $CS0[{l}^{\prime \prime \prime}] \gets \varnothing, {\mathbf{M}0}^{N}_{0}[{l}^{\prime \prime \prime}]  \gets \varnothing$ \;
    \For{$j \gets 0$ \To $Ls-1$}{
    	$(CS[j][s], {\mathbf{M}}^{N}_{0}[j][s])=CSSplit(CS1, CS[l^{\prime \prime}],{\mathbf{M}}^{N}_{0}[l^{\prime \prime}],Ls)$ \;
    $CS0[{l}^{\prime \prime \prime}]=CS0[{l}^{\prime \prime \prime}] \cup CS[j][s]$ \;
    ${\mathbf{M}0}^{N}_{0}[{l}^{\prime \prime \prime}]={\mathbf{M}0}^{N}_{0}[{l}^{\prime \prime \prime}] \cup {\mathbf{M}}^{N}_{0}[j][s]$ \;
  	}
    $(CS[l^{\prime \prime}], {\mathbf{M}}^{N}_{0}[l^{\prime \prime}],Ls)=CSPruning(CS1,CS0[{l}^{\prime \prime \prime}],{\mathbf{M}0}^{N}_{0}[{l}^{\prime \prime \prime}],Lc)$ \;
    $PSCS[i]=CS[0],i \gets  i+1$ \;
  }

\For{$i \gets |CS1|$ \To $|CS1|+|CS2|-1$}{
    $CS0[{l}^{\prime \prime \prime}] \gets \varnothing, {\mathbf{M}0}^{N}_{0}[{l}^{\prime \prime \prime}]  \gets \varnothing$ \;
    \For{$j \gets 0$ \To $Ls-1$}{
    	$(CS[j][s], {\mathbf{M}}^{N}_{0}[j][s])=CSSplit(CS2, CS[l^{\prime \prime}],{\mathbf{M}}^{N}_{0}[l^{\prime \prime}],Ls)$ \;
    $CS0[{l}^{\prime \prime \prime}]=CS0[{l}^{\prime \prime \prime}] \cup CS[j][s]$ \;
    ${\mathbf{M}0}^{N}_{0}[{l}^{\prime \prime \prime}]={\mathbf{M}0}^{N}_{0}[{l}^{\prime \prime \prime}] \cup {\mathbf{M}}^{N}_{0}[j][s]$ \;
  	}
    $(CS[l^{\prime \prime}], {\mathbf{M}}^{N}_{0}[l^{\prime \prime}],Ls)=CSPruning(CS2,CS0[{l}^{\prime \prime \prime}],{\mathbf{M}0}^{N}_{0}[{l}^{\prime \prime \prime}],Lc)$ \;
    $PSCS[i]=CS[0],i \gets  i+1$ \;
  }
\Return{$PSCS[l]$}
\end{algorithm}

The PSCS based construction is described in Algorithm 3, where $N$ is the code length, $K$ is the number of information bits, $\mathbf{g}$ is the convolutional polynomial, $\mathcal{A}$ is the information set, $L$ is the list size of SCL decoding, $Lc$ is the list size of PSCS, and $PSCS[l]$ is the obtained PSCS. 
The $SCLGetWS()$ function in line 1 uses SCL decoding to get partial weight spectrum of $\mathcal{A}$, $L$ source words estimates ${\hat{v}}^{N-1}_{0}[{l}^{\prime}]$ and codewords estimates ${\hat{x}}^{N-1}_{0}[{l}^{\prime}]$. 
${z}[{l}^{\prime}]$ is one of the weights corresponding to ${\hat{x}}^{N-1}_{0}[{l}^{\prime}]$, where $0\leqslant l^{\prime}<L$. 
The $getCS()$ function in line 2 makes decision between $CS1$ and $CS2$ based on $\mathcal{A}$, where $CS1$ is the set of all indexes for $s(i)=r$ in $\mathcal{A}$. 
If $\begin{matrix} K=\sum^{n}_{q=r} {\dbinom{n}{q}} \end{matrix}$, then $CS2$ is null; Otherwise $CS2$ is the set of all indexes of $s(i)=r+1$ in $\mathcal{A}$. 
Where $CS1 \cup CS2=CPCS$. 
The $CSRootInit()$ function in line 3 makes $CS[l^{\prime \prime}]=CS[l^{\prime \prime}] \cup {CS1}_{l^{\prime \prime}}$ and updates the LS metric ${\mathbf{M}}^{N}_{0}[l^{\prime \prime}]$. 
During LS metric ${\mathbf{M}}^{N}_{0}[l^{\prime \prime}]$ updating, if the bit of ${\hat{v}}^{N-1}_{0}[{l}^{\prime}]$ whose index is ${CS1}_{{l}^{\prime}}$ is 1, the $z[{l}^{\prime}]$-th element ${\mathbf{M}}_{z[{l}^{\prime}]}[l^{\prime \prime}]$ in ${\mathbf{M}}^{N}_{0}[l^{\prime \prime}]$ is increased by 1. 
$CSRootInit()$ function to initializes root node of the PSCS tree, and the root node size is $|CS1|$. 
For PSCS construction, the selection of this critical sets becomes more reliable if the LS metric is larger. 
After each sort operation is finished, the LS metric is cleared to zero. 
The $CSSplit()$ function in lines 8 and 18 splits respectively into $|CS1|-|CS[j]|$ and $|CS2|-|CS[j]|$ child nodes at the $j$-th parent node of the PSCS tree and updates the LS metric in the list. 
The $CSPruning()$ function in lines 12 and 22 keeps only one of the PSCS in the list. 
After the above steps, the $Ls$ child nodes are remained. 
If $Ls>Lc$, the maximum $Lc$ child nodes with LS metric are picked out, and the child nodes are sorted from largest to smallest according to LS metric. 
Finally, the obtained $PSCS[l]$ consist of $|CS1|+|CS2|$ PSCS. 

\begin{figure}[htb]
\centering
\includegraphics[width=3.5in]{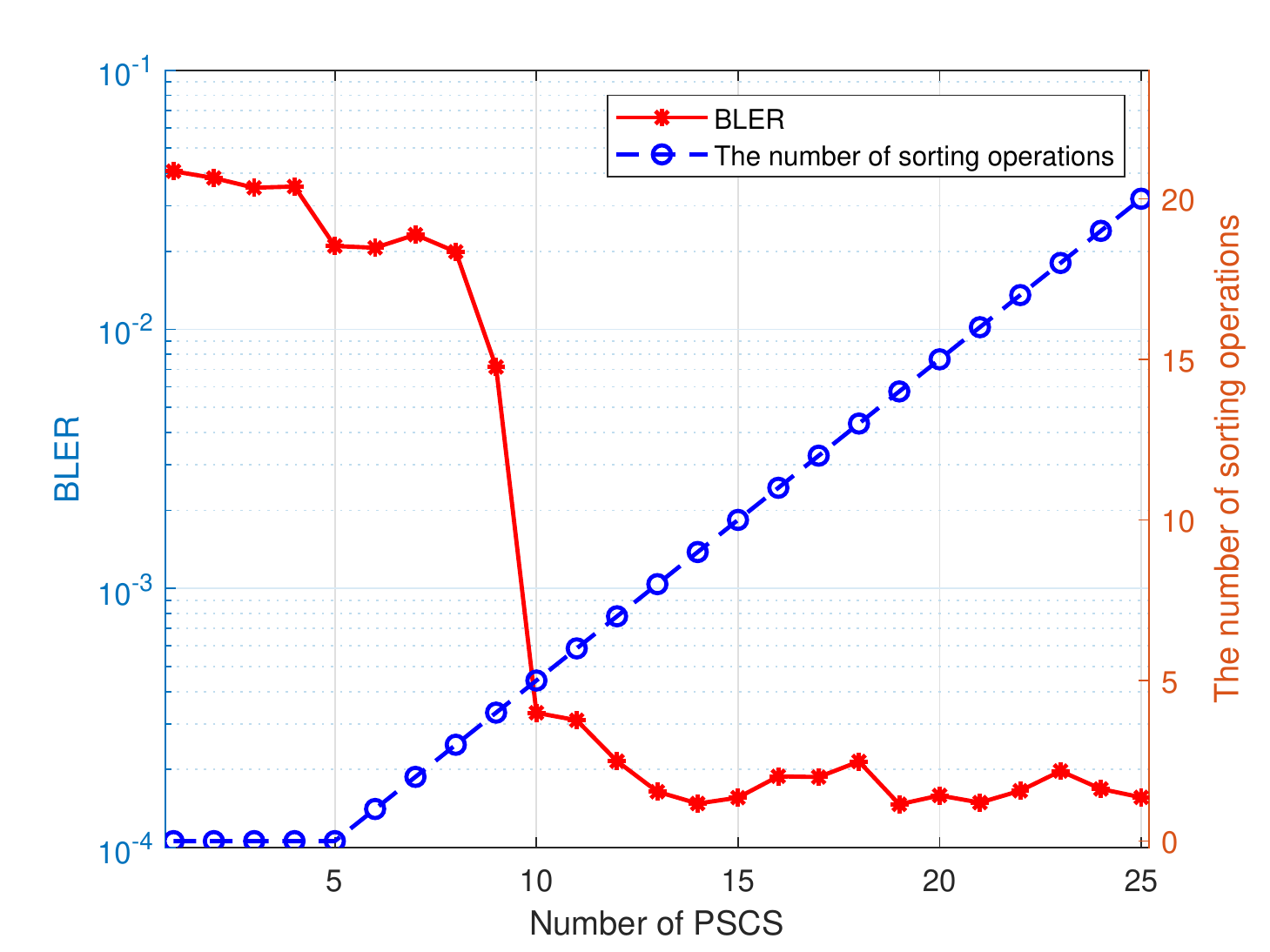}
\caption{BLER performance and the number of sorting operations for SCL decoding at 4dB for PAC(64,32) codes with different number of PSCS.}
\label{fig-11}
\end{figure}

During simulation, the parameters of PSCS construction include $L=20000$ and $Lc=400$. 
Figure 10 shows the BLER performance and number of sort operations for SCL decoding at 3dB for PAC(128,64) codes with different numbers of PSCS, and the list size $L=128$. 
It can be seen that there is negligible difference on BLER performance of SCL(PSCS) decoding when the number of PSCS is from 29 to 35, while number of sort operations is from 22 to 28 respectively. 
Similarly, Figure 11 shows the BLER performance and number of sort operations for SCL decoding at 4dB for PAC(64,32) codes with different numbers of PSCS, and the list size $L=32$. 
There is negligible difference on BLER performance of SCL(PSCS) decoding when the number of PSCS is from 12 to 28, while number of sort operations is from 7 to 20 respectively. 
When there is negligible difference on BLER performance of SCL(PSCS) decoding and SCL(CPSCS)decoding, we can find that number of sort oprations is further reduced from 28 to 22 in Figures 8 and Figures 10, and from 20 to 7 in Figures 9 and Figures 11.

\section{Simulation Results}
All simulation results presented in this section are based on binary input additive Gaussian White noise (BI-AWGN) channels and BPSK modulation. 
Threshold spacing $\Delta=2$ for Fano decoding, and Fano metric is caculated by the metric function in \cite{ref23}. 
Defining a performance deterioration which is tolerable to the system requirements is ${P}_{tol}=1.0\times {10}^{-5}$ for Pruned SCL decoding \cite{ref16}, and the PSCL decoding threshold ${m}_{T}=-10$ \cite{ref18}. 
If the convolution polynomial is not specified, the default is $\mathbf{g1}=\{1,1,0,1,0,0,0,1,0,0,1\}$ (3211 in octal notation). 

\begin{table*}[t]
\caption{Rate profiles of the LS based construction with retained duplicate path
\label{tab:table2}}
\centering
\begin{tabular}{c c c c c c}
\hline
$(N,K)$ & $\mathbf{g}$ & $L$ & $Lg$ & ${E}_{b}/{N}_{0}$(dB) & Rate Profile($\mathbf{\alpha}$) \\
\hline
\multirow{1}{*}{$(256,128)$} & \multirow{1}{*}{$\mathbf{g2}=0o133$} & 40000 & 400 & 3.2 & 00000005001511770015155717575FFF0015155717577F7F171737FF177F7FFF \\
\hline
\end{tabular}
\end{table*}

\subsection{List-Search rate-profiles}

\begin{figure}[t]
\centering
\includegraphics[width=3.5in]{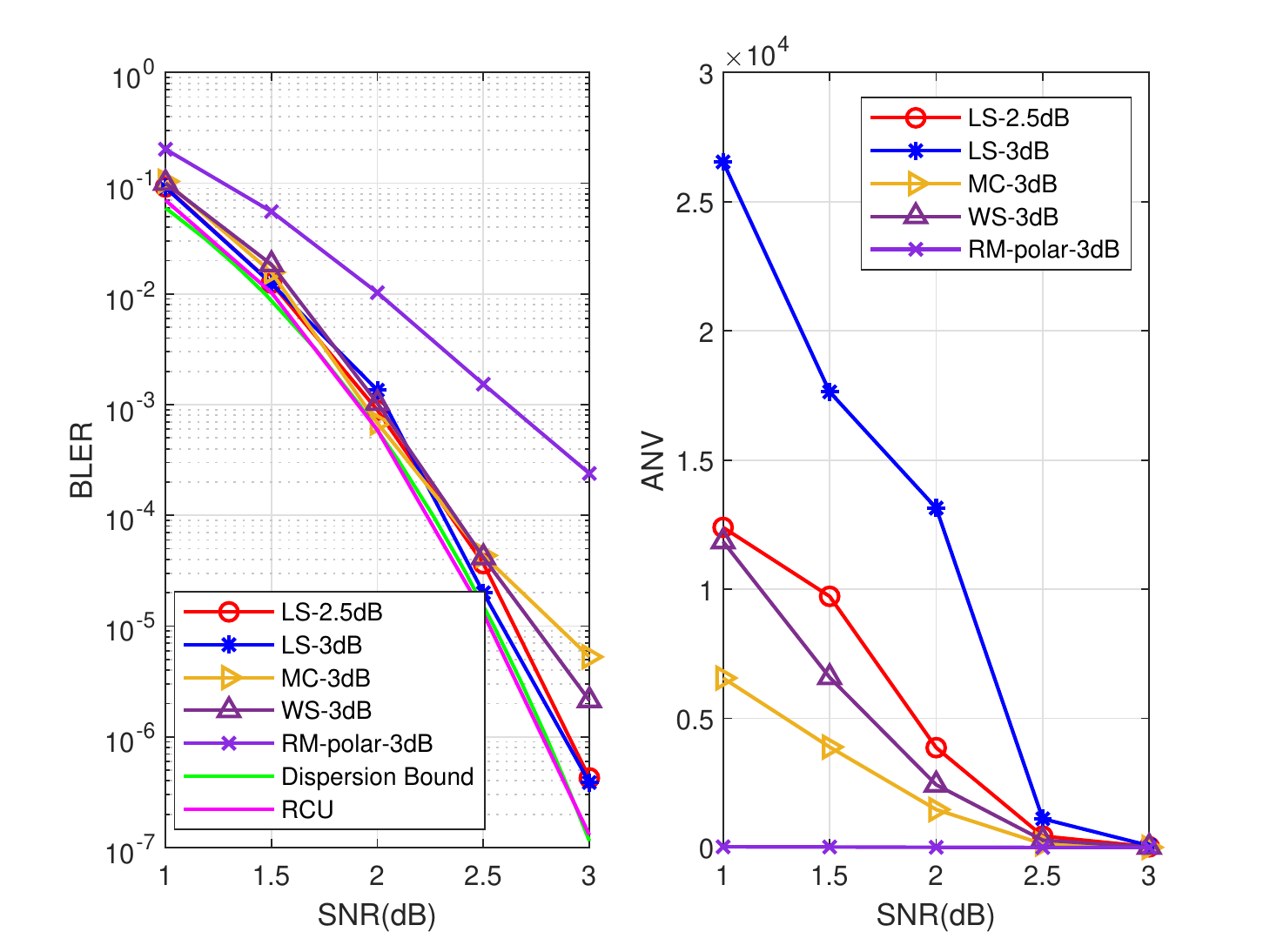}
\caption{BLER performance and ANV performance of Fano decoding for PAC(256,128) codes.}
\label{fig-12}
\end{figure}

Average number of visits (ANV) defined in \cite{ref12} were used to evaluate the decoding complexity of Fano decoding for different rate-profiles. 
In addition, the dispersion bound \cite{ref9} and the random-coding union (RCU) bound \cite{ref9} have been drawn. 
Figure 12 shows the BLER performance and ANV performance of Fano decoding for PAC(256,128) codes with a convolution polynomial of $\mathbf{g1}$. 
The BLER performance of LS-2.5dB rate-profiles and LS-3dB rate-profiles is very close to the dispersion bound, and better than the BLER performance of MC-3dB rate-profiles and WS-3dB rate-profiles at high SNR. 
ANV performance of LS-2.5dB rate-profiles is superior to that of LS-3dB rate-profiles without visible degradation in BLER performance. 

Figure 13 shows the BLER performance and ANV performance of Fano decoding for PAC(256,128) codes with two different convolutional polynomials. 
When the convolution polynomial is $\mathbf{g2}$ in Table I, the BLER performance of LS-2.5dB rate-profiles and LS-3.2dB rate-profiles deviates from the dispersion bound. 
Table II shows the rate-profiles obtained by retained duplicate path (RDP) during path pruning of LS construction. 
While the ANV performance of LS-3.2dB rate-profiles is close to that of LS-3.2dB(RDP) rate-profiles, the BLER performance of LS-3.2dB(RDP) rate-profiles is close to the dispersion bound, and superior to that of LS-3.2dB rate-profiles in Figure 13. 
When the convolution polynomial is $\mathbf{g3}$ (a special case of PAC codes: polar codes) in Table I, LS-3.2dB rate-profiles has approximately 0.2dB gain over RM-polar-3dB rate-profiles at $BLER={10}^{-3}$. 

\begin{figure}[t]
\centering
\includegraphics[width=3.5in]{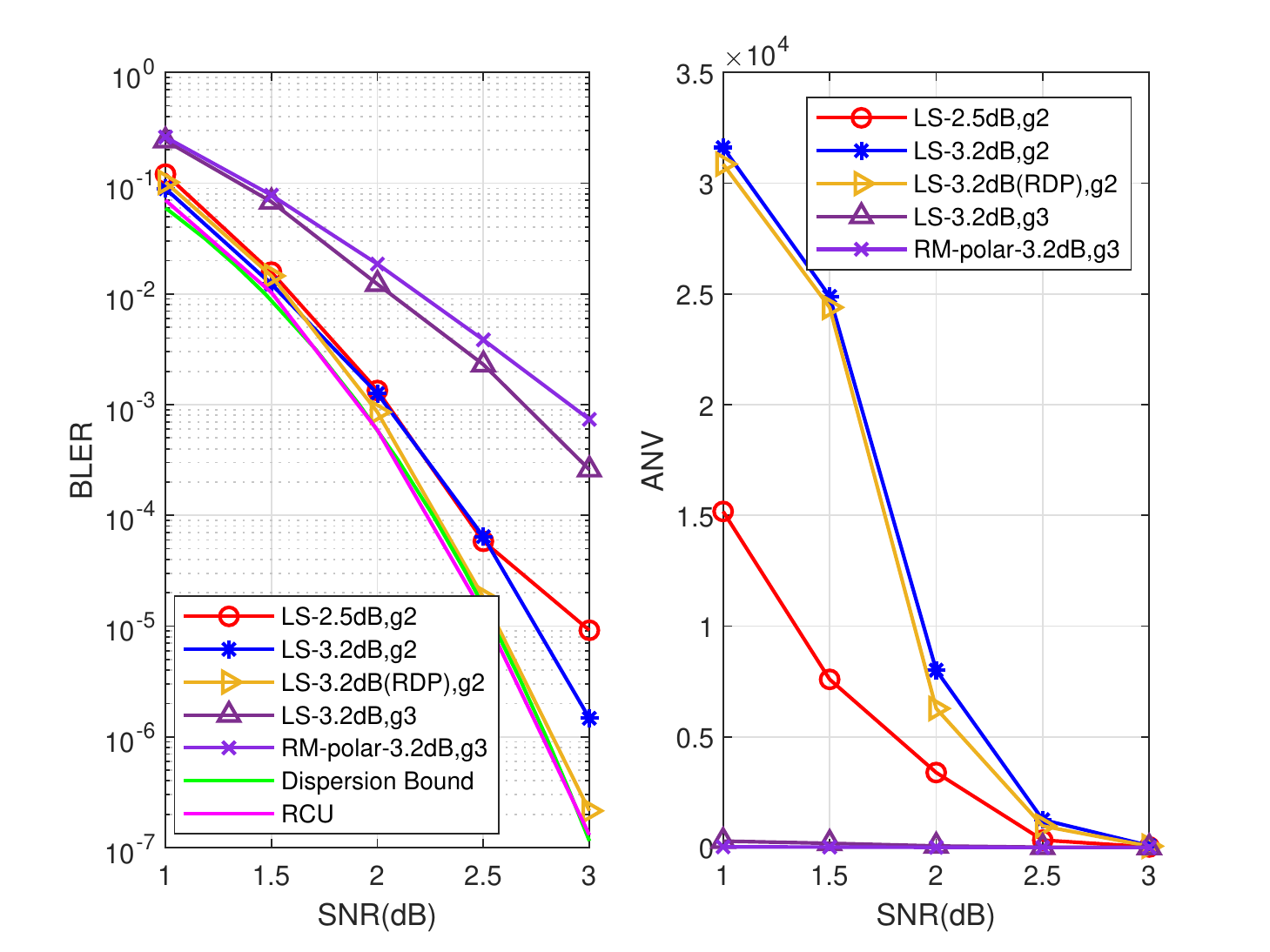}
\caption{BLER performance and ANV performance of Fano decoding for PAC(256,128) codes  with different convolutional polynomials.}
\label{fig-13}
\end{figure}

\begin{figure}[htb]
\centering
\includegraphics[width=3.5in]{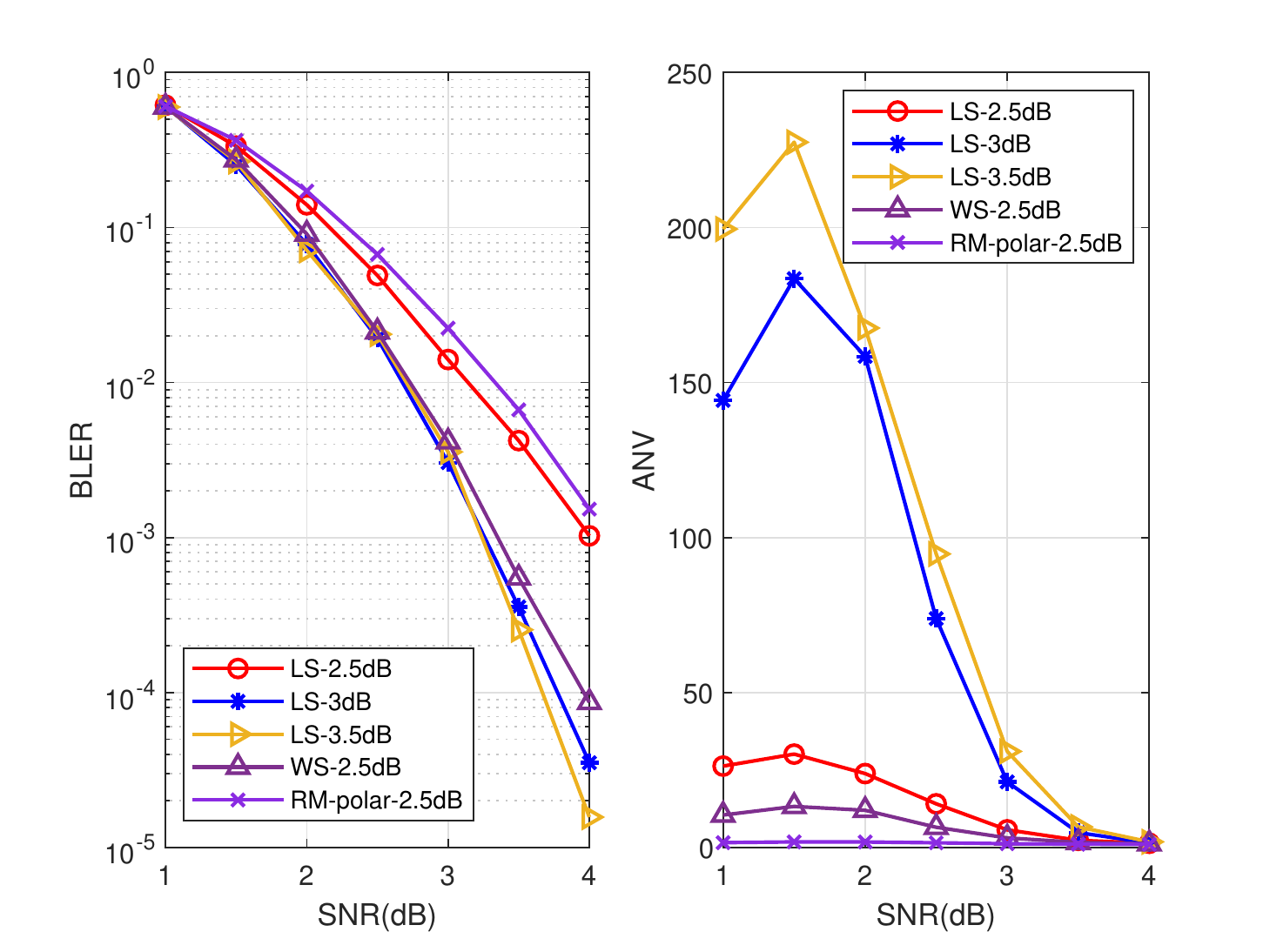}
\caption{BLER performance and ANV performance of Fano decoding for PAC(128,85) codes.}
\label{fig-14}
\end{figure}

In Figure 14, the WS-2.5dB rate-profiles are the most optimal rate-profiles obtained in \cite{ref13}, and the BLER performance of WS-2.5dB rate-profiles is better than that of LS-2.5dB rate-profiles. 
However, LS-3.0dB rate-profiles and LS-3.5dB rate-profiles have better BLER performance than WS-2.5dB rate-profiles, especially in high SNR region. 
In Figure 15, the three LS rate-profiles have better BLER performance than WS-2.5dB rate-profiles, and the LS-3.5dB rate-profiles gains approximately 0.4dB over the WS-2.5dB rate-profiles when $BLER={10}^{-3}$. 

\begin{figure}[t]
\centering
\includegraphics[width=3.5in]{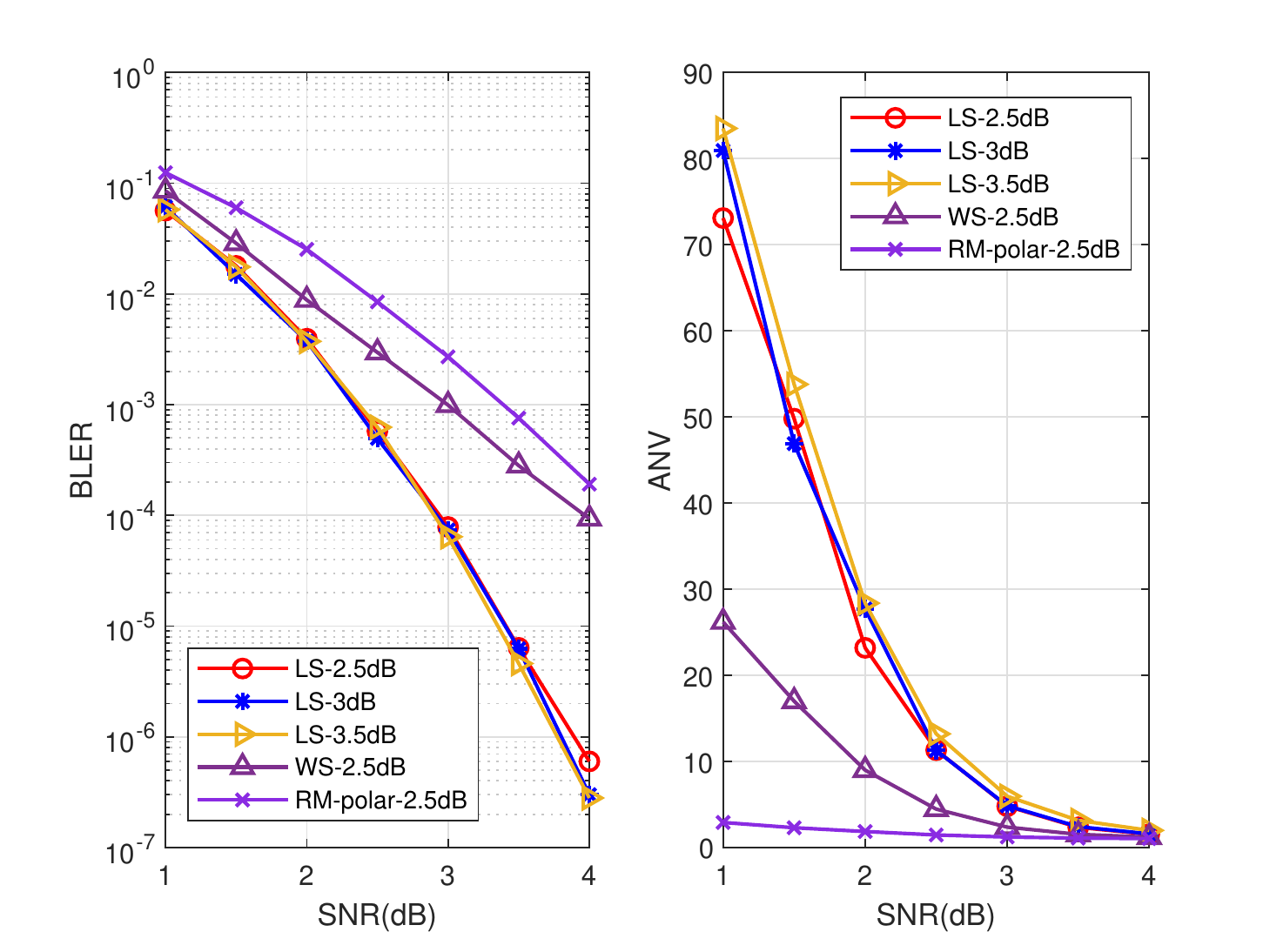}
\caption{ BLER performance and ANV performance of Fano decoding for PAC(128,42) codes.}
\label{fig-15}
\end{figure}

\begin{figure}[htb]
\centering
\includegraphics[width=3.5in]{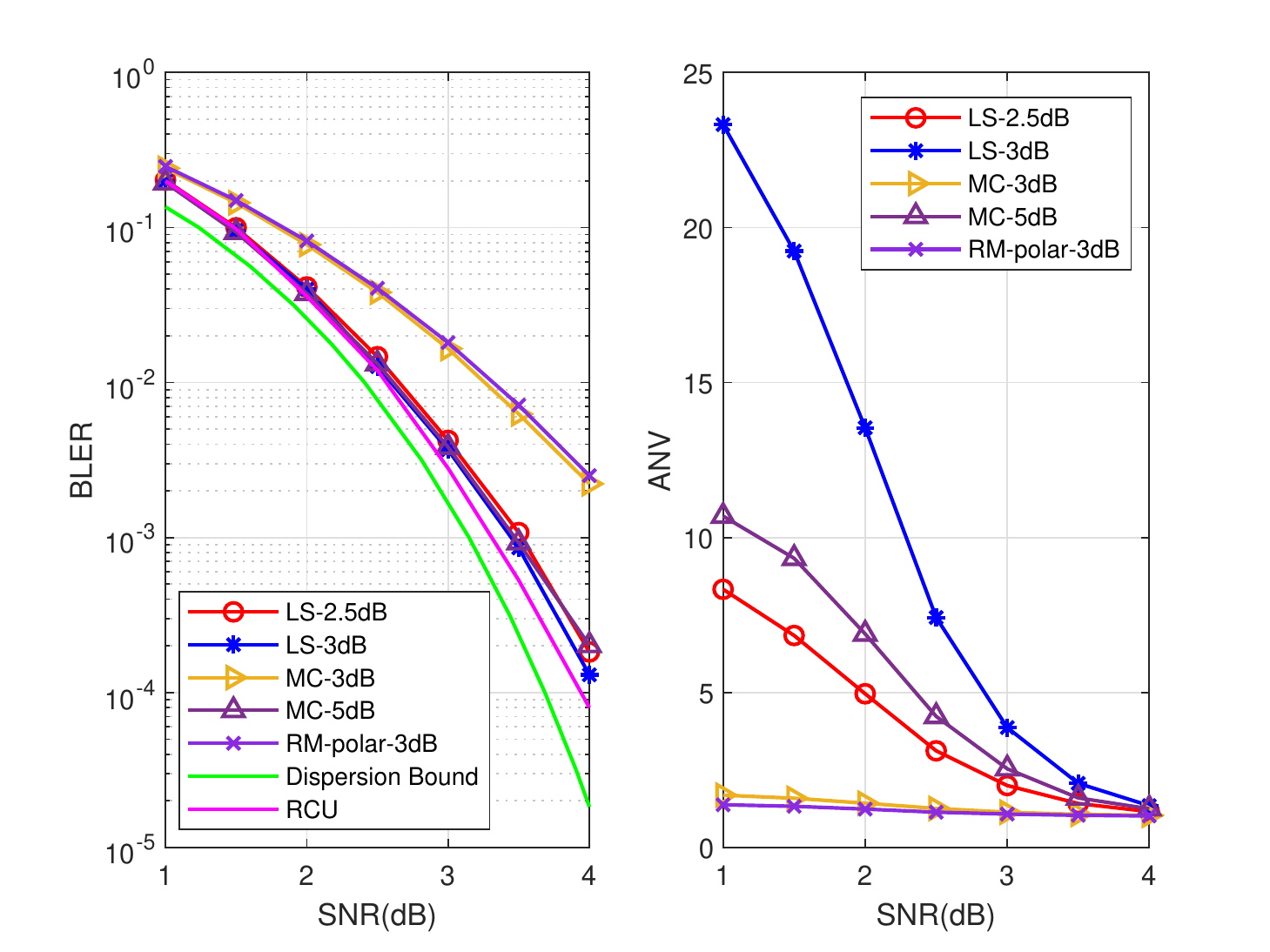}
\caption{ BLER performance and ANV performance of Fano decoding for PAC(64,32) codes.}
\label{fig-16}
\end{figure}

In Figure 16, when there is negligible difference on BLER performance of LS-2.5dB rate-profiles and MC-5dB rate-profiles, the ANV performance of LS-2.5dB rate-profiles is superior to that of MC-5dB rate-profiles. 
The BLER performance of LS-3dB rate-profiles is better than that of MC-3dB rate-profiles, MC-5dB rate-profiles. 
LS-3dB rate-profiles has approximately 0.3dB gain over the MC-3dB rate-profiles at $BLER={10}^{-2}$. 
The BLER performance of LS-2.5dB rate-profiles or LS-3dB rate-profiles is close to the RCU bound. 

The ANV values of PAC(64,32) codes are significantly smaller than PAC(256,128) codes, so SCL decoding for PAC(64,32) codes with a small list size is feasible \cite{ref12}. 
Figure 17 shows the BLER performance and ANV performance of SCL decoding for PAC(64,32) codes. 
When the list size is 32 or 8, the performance of the LS-2.5dB rate-profiles is better than that of the MC-5dB rate-profiles. 
As the list size increases, the performance of SCL decoding for PAC(64,32) codes approaches the RCU bound. 
The BLER performance of LS-2.5dB rate-profiles with list size of 8 is superior to that of RM-polar-3dB rate-profiles. 
We can conclude that that the constructed PAC codes by LS have good BLER performance of SCL decoding. 

\begin{figure}[htb]
\centering
\includegraphics[width=3.5in]{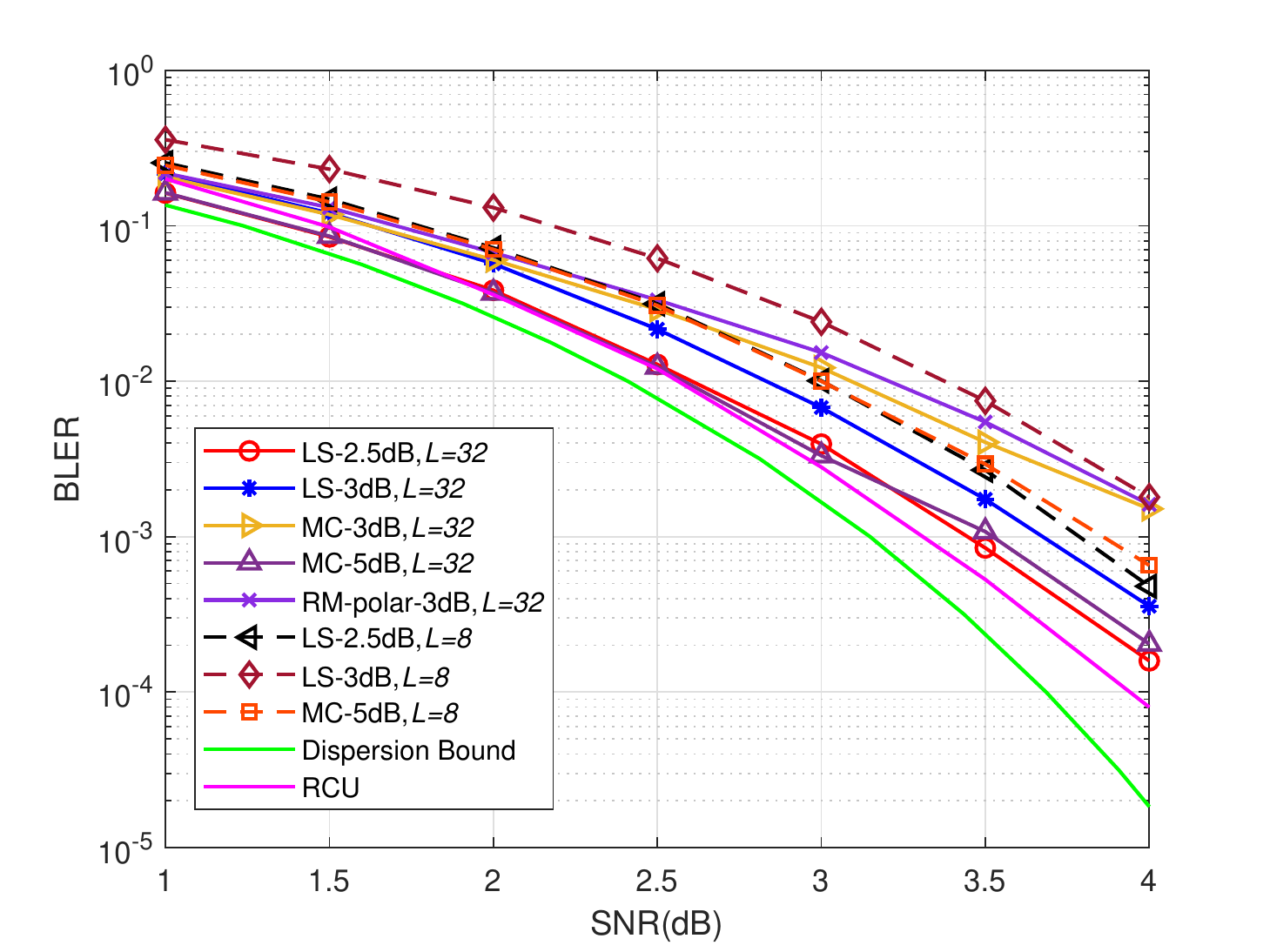}
\caption{ BLER performance and ANV performance of SCL decoding for PAC(64,32) codes.}
\label{fig-17}
\end{figure}

From Figure 12 to 16, we can conclude that that LS rate-profiles has better BLER performance than MC rate-profiles or WS rate-profiles during Fano decoding for PAC codes. 
Moreover, the BLER performance of LS rate-profiles can reach the theoretical bound in high SNR region. 

\subsection{The Path-Splitting Critical Sets}
Figure 18 shows the BLER performance and the number of sort operations  during various SCL-type decoding for PAC(128,64) codes before and after PSCS construction when the list size is 128. 
Figure 18(a) shows that the constructed codes with PSCS can acheive BLER performance of each SCL-type decoding very close to those without PSCS, but in Figure 18(b), the number of sort operations during list decoding is significantly reduced after using PSCS construction method. 
Similarly, Figure 19 shows the BLER performance and the number of sort operations of various SCL-type decoding for PAC(64,32) codes when the list size is 32. 
It can be seen that the BLER performance of SCL decoding, Pruned SCL decoding, or PSCL decoding is very close for the constructed PAC(64,32) codes with PSCS or without PSCS, but the number of sorting operations of each SCL-type decoding is significantly reduced after using PSCS based construction method.

\section{Conclusion}
\begin{figure*}[htbp]
\centering
\subfloat[BLER performance.]{
\begin{minipage}[t]{0.48\linewidth}
\centering
\includegraphics[width=3.5in]{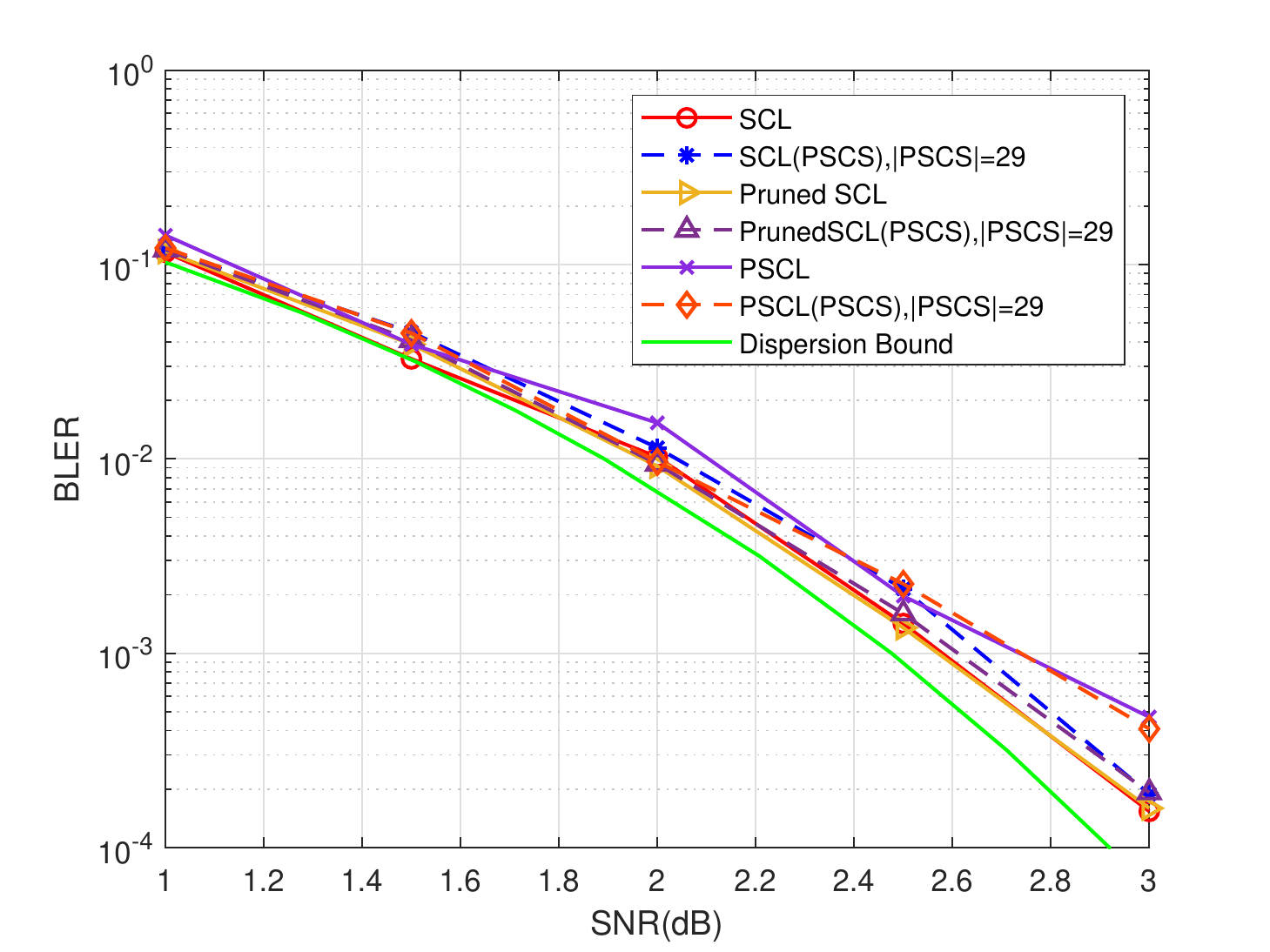}
\end{minipage}%
}%
\subfloat[The number of sorting operations.]{
\begin{minipage}[t]{0.48\linewidth}
\centering
\includegraphics[width=3.5in]{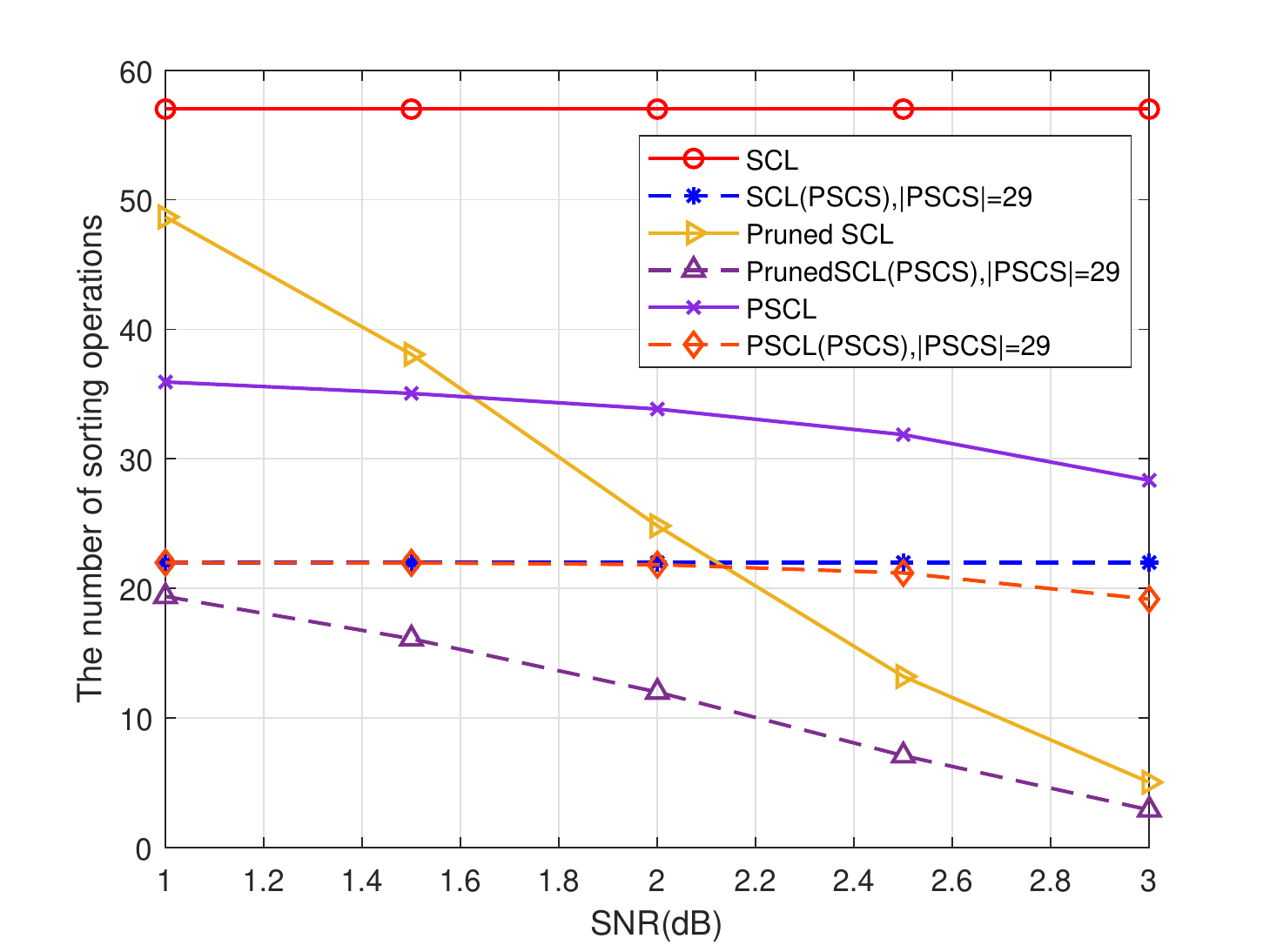}
\end{minipage}%
}%
\centering
\caption{The BLER performance and the number of sorting operations of various SCL-type decoding for PAC(128,64) codes.}
\end{figure*}

\begin{figure*}[htbp]
\centering
\subfloat[BLER performance.]{
\begin{minipage}[t]{0.48\linewidth}
\centering
\includegraphics[width=3.5in]{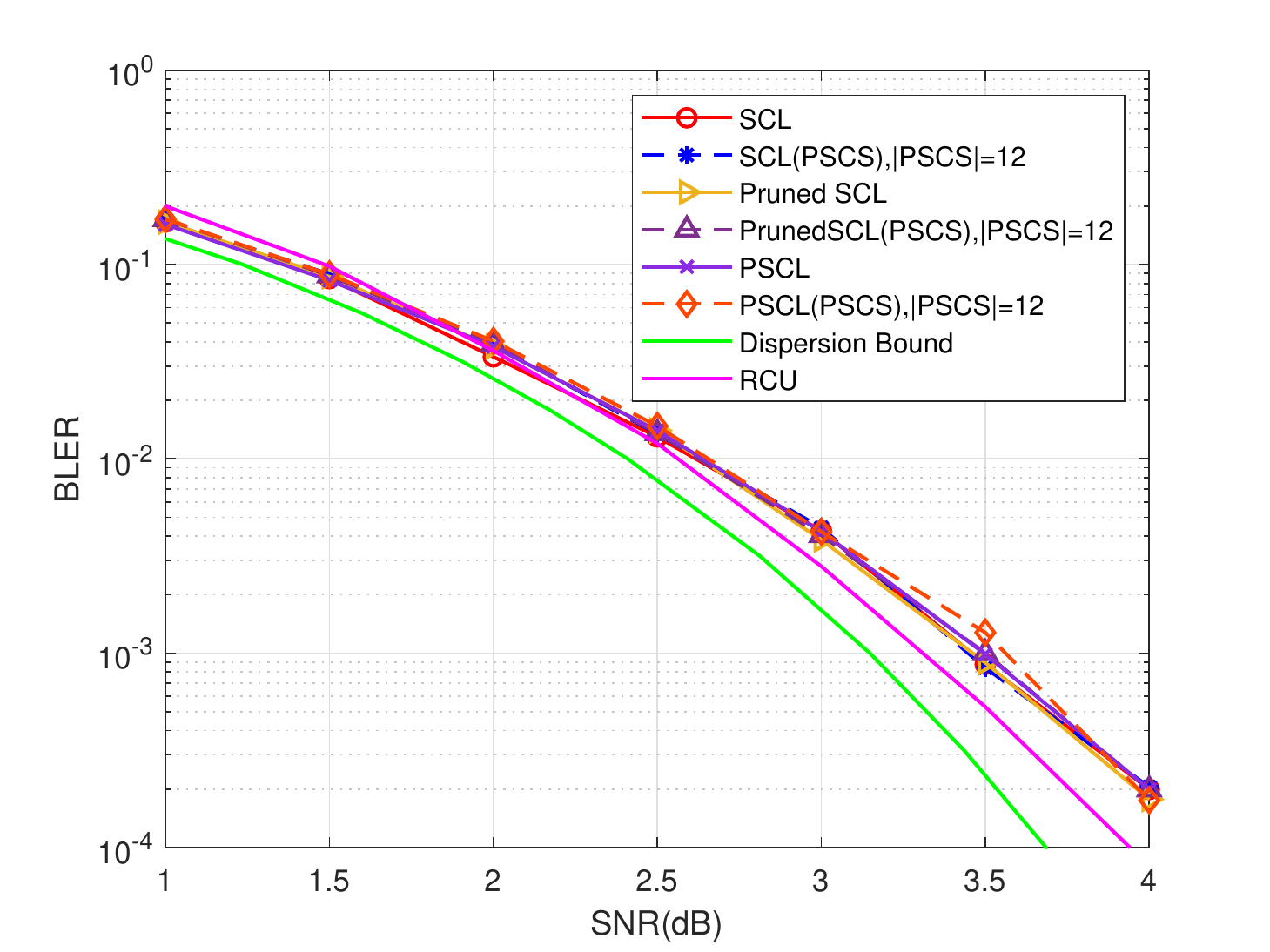}
\end{minipage}%
}%
\subfloat[The number of sorting operations.]{
\begin{minipage}[t]{0.48\linewidth}
\centering
\includegraphics[width=3.5in]{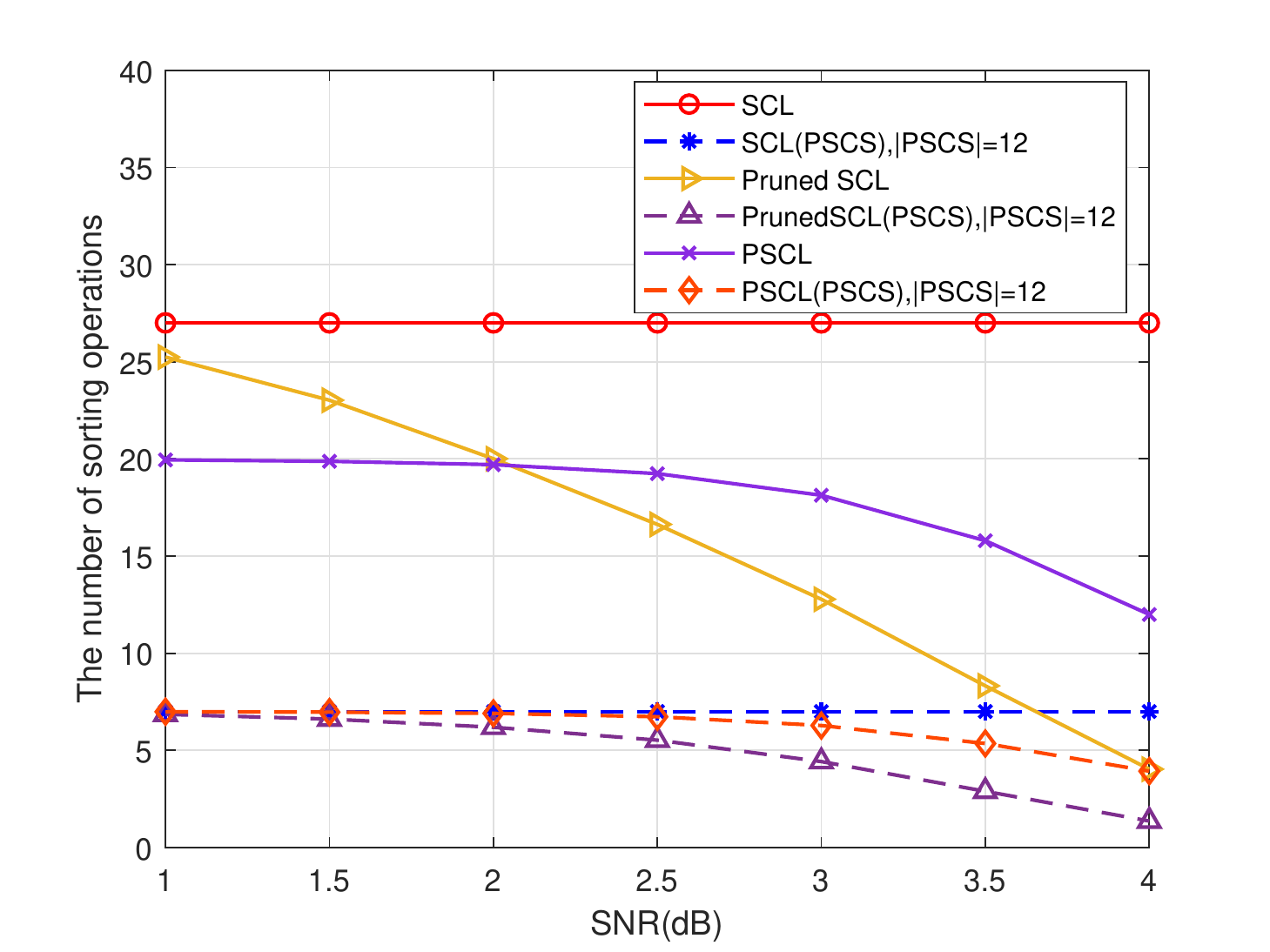}
\end{minipage}%
}%
\centering
\caption{The BLER performance and the number of sorting operations of various SCL-type decoding for PAC(64,32) codes.}
\end{figure*}

In this paper, we proposed a LS based construction for PAC codes of any code length and any code rate. 
The decoding performance of the constructed codes is close to the theoretical bound. 
This method analyzes the influence of weight spectrum on the BLER performance of ML decoding, and then gets the principle of the constructed codes with weight maximization and the number of minimum weight codewords. 
To get the constructed codes with a limited average computational complexity during sequential decoding, this method need to remove the child nodes that do not satisfy one specific condition. 
The simulation results show that LS rate-profiles for PAC(256,128) codes and PAC(64,32) codes can approach the theoretical bound at high SNR, compared with MC rate-profiles and WS rate-profiles. 
In addition, for PAC(256,128) codes with a convolution polynomial of $\mathbf{g3}$ (a special case of PAC codes: polar codes), the LS rate-profiles has a gain of about 0.2dB over the RM-polar rate-profiles at $BLER={10}^{-2}$.

Moreover, this paper proposes the PSCS based construction of PAC codes for RM rate-profiles and generalization of RM rate-profiles. 
The simulation results show that this method can significantly reduce the number of sort operations during SCL decoding, Pruned SCL decoding and PSCL decoding with no visible degradation in the BLER performance of the above SCL-type decoding for the constructed PAC codes with PSCS.

\balance



\end{document}